\documentclass[twoside,twocolumn,english]{revtex4-2}
\usepackage{amsmath}
\usepackage{mathptmx}
\usepackage{newtxmath}

\usepackage[T1]{fontenc}
\usepackage[utf8]{luainputenc}
\usepackage{color}
\usepackage{babel}
\usepackage{mathtools}
\usepackage{dsfont}
\usepackage{graphicx}
\usepackage[unicode=true,
 bookmarks=true,bookmarksnumbered=false,bookmarksopen=false,
 breaklinks=false,pdfborder={0 0 1},backref=false,colorlinks=true]
 {hyperref}
\hypersetup{linkcolor=magenta,urlcolor=marineblue,citecolor=blue,pdfstartview={FitH},hyperfootnotes=false,urlcolor=blue}

\makeatletter

\newcommand{\lyxdot}{.}

\usepackage{babel}
\usepackage{braket}

\makeatother

\begin{document}
\title{Suppressing information storage in a structured thermal bath: Objectivity and non-Markovianity }
\author{Wilson S. Martins}
\email{wilson.santana.martins@usp.br}
\affiliation{Instituto de F\'{i}sica de S\~{a}o Carlos, Universidade de S\~{a}o Paulo, CP 369, 13560-970, S\~{a}o Carlos, SP, Brazil}
\author{Diogo O. Soares-Pinto}
\email{dosp@usp.br}

\affiliation{Instituto de F\'{i}sica de S\~{a}o Carlos, Universidade de S\~{a}o Paulo, CP 369, 13560-970, S\~{a}o Carlos, SP, Brazil}
\begin{abstract}
Quantum systems interacting with environments tend to have their information
lost from transmission through the correlations generated with the
degrees of freedom of the environment. For situations where we have
non- Markovian environments, the information contained in the environment
may return to the system and such effect can be captured by non-Markovian
witnesses. In the present work, we use a central qubit coupled to
a spin chain with Ising interactions subject to a magnetic field,
i.e., a central spin model, and solve the exact dynamics of a system
subject to dephasing dynamics via Kraus operators. For such we use
two witnesses to analyze the presence of non-Markovianity: the BLP
trace-based measure and the conditional past-future correlator (CPF).
Furthermore, we see how such behavior suppresses the classic plateau
in Partial Information Plot (PIP) from the paradigm of quantum Darwinism,
as well as objective information from Spectrum Broadcast Structure
(SBS). In addition to the system point of view, we show the impossibility
of encoding accessible information for measurement in the environment
for any model limit. 
\end{abstract}
\maketitle

\section{\label{sec1}Introduction}

A quantum system in contact with an environment, i.e., an open quantum
system, is subject to decoherence, which means that we lose states
of quantum superposition as a result of the classical uncertainty
created by the persistent loss of information to the environment.
What we have, then, is that the disturbance caused by the environment
promotes the loss of quantumness of the systems that would preserve
this nature, in principle, when isolated \citep{sep-qm-decoherence,Schlosshauer_2005}.
This situation gives us a crucial overview of the experiments, which
will have to be mediated taking into account the sensitivity of states
to the environment, hence the importance of studying open quan- tum
systems. An important class of such open systems is those coupled
to finite baths with finite temperatures, where the transmission of
system-environment information and vice versa becomes non-trivial,
introducing, for example, situations where information previously
lost to the environment returns for the system of interest.

The characterization of information return from the environment to
a given system can be analyzed in the light of recoherence, indicated,
for example, by non-Markovianity. When we are dealing with classical
systems, the system's Markovianity reflects the divisibility of the
transition maps of a stochastic process \citep{vacchini2011markovianity,rivas2014quantum,li2018concepts}.
This implies that in a Markovian system the states defined at some
point in time do not depend on previous states, that is, there is
the emergence of the memoryless property. Notably, this intuitive
notion allows for some new ways to characterize non-Markovianity in
open quantum systems, where we can, for example, characterize such
systems using state distinguishability based on the distance of distinct
states in the state space, such as the Breuer-Laine-Piilo (BLP) measure,
proposed by \citep{laine2010measure}, or by the correlation between
states measured at different points on the timeline, as described
by the conditional past-future correlation (CPF) \citep{Budini_2019,Budini_2018}.
For both previous programs, we are using the observation of what is
happening in the quantum system of interest to identify feedback of
information. We can, of course, take another point of view, where
the deposit of information in the environment and its accessibility
are important. For this analysis, we can use, respectively, the paradigms
of quantum Darwinism \citep{Zurek_2009,Zurek2014,Zurek2011,zurek2007relative,touil2021eavesdropping},
and the Spectrum Broadcast Structure (SBS) \citep{Horodecki2015,Korbicz2020,le2019strong},
which answer questions about encoded and accessible information for
knowledge from the perspective of observers for whom the environment
is accessible.

\begin{figure}
\includegraphics[scale=0.4]{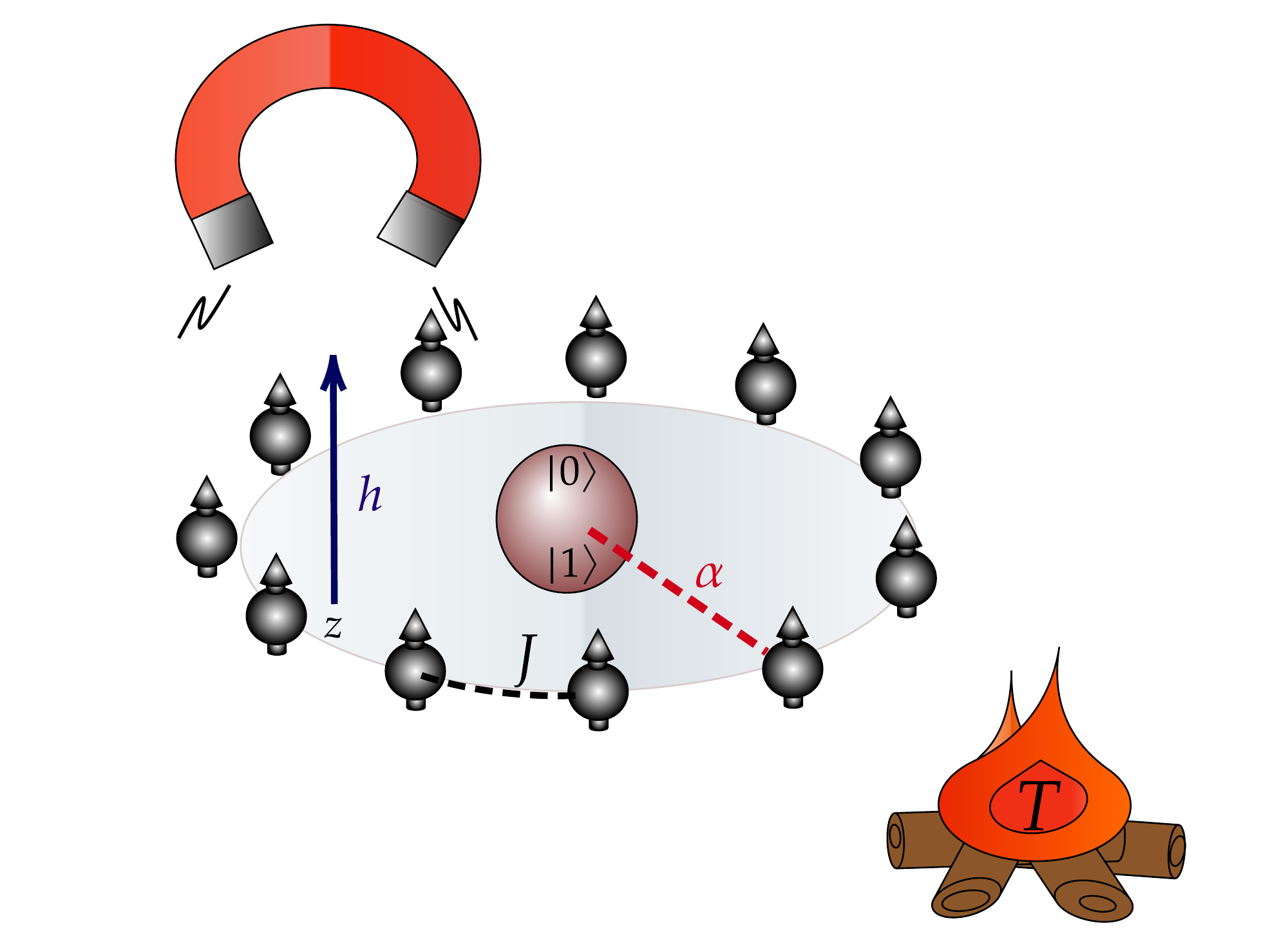}

\caption{(Color online) Here, we use a central qubit coupled to a thermal bath
structured by spins with ferromagnetic interactions between the first
neighbors (with intensity $J>0$), also subject to the action of a
magnetic field $h$ in the direction $\hat{z}$. The bath is in thermal
equilibrium at the temperature $T=\frac{1}{\beta}$. Along the paper
we consider $k_{B}=\hbar=1$. \label{model}}
\end{figure}

In this paper, we consider these physical descriptions to study and
characterize a platform of an open quantum system based on a qubit
coupled to a thermal bath with finite temperature, containing spins
subjects to Ising-like ferromagnetic interactions between the nearest-neighbors
and a magnetic field (Fig.\ref{model}). The present model is based
on the model used by Ref. \citep{Wei_2012}, in which they describe
the interrelations between the appearance of Lee-Yang zeros in the
fugacity plan and the critical times of the coherence function, i.e,
the zeros of the coherence function in the time domain \citep{PhysRevLett.114.010601}.
Our fundamental purpose is to relate the non-Markovian behavior (including
its witnesses and measurements) as well as the storage of information
in the environment (which we will indistinctly call bath to the throughout
the text), aiming at an overview from the point of view of the system
and the environment. For this, we will investigate the suppression
of both quantum Darwinism and SBS..

The structure of the paper is arranged as follows. In Sec.\ref{sec2}
we construct a description of the system and the environment, the
results about its dynamics using Kraus representation, and the closed
relation between the coherence function and the Lee-Yang zeros. In
Sec.\ref{sec3} we present two non-Markovianity witnesses and its
respective plots to show the limits of non-Markovianity with respect
to the interacting parameters - magnetic field and Ising coupling
- and temperature. The Sec.\ref{sec4} contains the informational
description of the problem using the key ideas of quantum Darwinism
and SBS, explaining the insufficiency of first to explain the objectivity,
the importance of the distinguishability between the broadcast states
to the emergence of objectivity, and the results for the present model,
while the Sec.\ref{sec5} presents the conclusions.

\section{\label{sec2}Description of the model}

Here we will defining the notation and the structure of the representative
Hilbert spaces. Denoting $\mathcal{H_{S}}$ the Hilbert space representative
of the states of the central qubit and by $\mathcal{H_{B}}$ the Hilbert
space of the spin bath states. The total system is then given by the
tensor product space $\mathcal{H_{SB}}=\mathcal{H_{S}}\otimes\mathcal{H_{B}}$.
For the Hilbert state with respect to the central qubit, we take a
\emph{2-dimensional} complex space in the computational basis, i.e.,
$\mathcal{H_{S}}=\mathrm{span}\left\{ \Ket{0},\Ket{1}\right\} \cong\mathbb{C}^{2}$,
and this states $\Ket{0},\Ket{1}$ are the eigenstates of the Pauli
matrix $\sigma^{z}$ correspondents to the eigenvalues $1$ and $-1$,
respectively.

Let us consider a Hilbert subspace that can be described too by a
\emph{2-dimensional }complex space, $\mathcal{H}_{k}=\mathrm{span}\{\Ket{0},\Ket{1}\}\cong\mathbb{C}^{2}$.
This space is the subspace descriptive of the $k$-th spin site in
the bath. As one can see, the bath can be construct as the tensor
prooduct of subspace corresponding to each site 
\[
\mathcal{H_{B}}=\bigotimes_{k=1}^{N}\mathcal{H}_{k}\cong(\mathbb{C}^{2})^{\otimes N},
\]
and the notation $(\mathbb{C}^{2})^{\otimes N}$ means the composition
of $N$ \emph{2-dimensional} complex spaces.

To know how the environment acquires and record information about
the central system, we will use the \emph{environment point-of-view}.
Thereby, it is important to construct partial Hilbert spaces, that
will be tensorial compositions of such number (lower than $N$) of
subsystems. If one call this fractional Hilbert space $\mathcal{F}$,
it can be written as 
\[
\mathcal{F}=\mathcal{H}_{1}\otimes...\otimes\mathcal{H}_{fN}=\bigotimes_{k=1}^{fN}\mathcal{H}_{k},
\]
such that $fN\equiv\#\mathcal{F}\le N$, and $\#$ represents the
number of composed subsystems (or the cardinality of $\mathcal{F}$
with respect to $\mathcal{H}_{k}$).

Now, with each Hilbert space defined, we can propose that the total
Hamiltonian that describes system-environment is given by 
\begin{equation}
H=H_{S}+H_{B}+H_{SB},
\end{equation}
where $H_{S}=\frac{\omega\sigma^{z}}{2}$ is the Hamiltonian of the
system and $H_{B}$ is the Ising-like environment Hamiltonian of $N$
spin $1/2$ $z$-axis, described by 
\begin{equation}
H_{B}=-J\sum_{i=1}^{N}\sigma_{i}^{z}\otimes\sigma_{i+1}^{z}-h\sum_{i=1}^{N}\sigma_{i}^{z},
\end{equation}
that contains Pauli matrices $\sigma_{i}^{z}$ acting in each space
$\mathcal{H}_{i}$, $J$ corresponds to a nearest-neighbor coupling
between the spins and $h$, the magnetic field along $z$-axis affecting
the spin chain.

On the one hand, the initial state of the system will be considered
as a general qubit state in the Bloch sphere 
\begin{equation}
\Ket{\psi}=a\Ket{0}+b\Ket{1}\in\mathcal{H_{S}},\label{initial_state}
\end{equation}
where $a,b\in\mathbb{C}$ and $|a|^{2}+|b|^{2}=1$.

On the other hand, since this qubit is subject to dynamics due to
the interaction with a thermal bath at temperature $T$, we will consider
that such bath state is described by the Gibbs state 
\begin{equation}
\rho_{B}=\frac{e^{-\beta H_{B}}}{Z_{B}},\label{initial_bath}
\end{equation}
where $Z_{B}\coloneqq\mathrm{Tr}[e^{-\beta H_{B}}]$ the partition
function.

Such Gibbs state will gives us a distribution in the state space and
one can construct each microstate as 
\begin{equation}
\Ket{\boldsymbol{\chi}}\coloneqq\Ket{\chi_{1}...\chi_{N}}\equiv\Ket{\chi_{1}}\otimes...\otimes\Ket{\chi_{N}},
\end{equation}
that is the state that diagonalizes $H_{B}$, which $\Ket{\chi_{i}}\in\mathcal{H}_{i}$
is a eigenstate of the Pauli matrix $\sigma^{z}$ in computational
basis $\left\{ \Ket{0},\Ket{1}\right\} $, correspondent to the eigenvalues
$\sigma_{i}=\pm1$. Then, the diagonal Hamiltonian results in 
\begin{equation}
H_{B}\Ket{\boldsymbol{\chi}}=E(\boldsymbol{\chi})\Ket{\boldsymbol{\chi}},
\end{equation}
defines the configuration correspondent to the energy 
\begin{equation}
E(\boldsymbol{\chi})=-J\sum_{i\in\mathcal{A}}\sigma_{i}\sigma_{i+1}-h\sum_{i\in\mathcal{A}}\sigma_{i}.
\end{equation}
for some $\mathcal{A}\subseteq\mathcal{H_{B}}$. As we will be interested
in bath fractions, sometimes $\mathcal{A}$ will correspond to a subspace
smaller than the space corresponding to the bath $\mathcal{H_{B}}$.

The diagonalization of the Ising chain Hamiltonian allows us to re-write
bath density operator in the energy basis: 
\begin{equation}
\rho_{B}=\frac{e^{-\beta H_{B}}}{Z_{B}}=\frac{1}{Z_{B}}\sum_{\boldsymbol{\chi}}e^{-\beta E(\boldsymbol{\chi})}\Ket{\boldsymbol{\chi}}\Bra{\boldsymbol{\chi}}.\label{do_energy}
\end{equation}
This system presents ground states obtained from $\lim_{\beta\rightarrow\infty}\rho_{B}$,
in which for $h=0$ emerges a $\mathbb{Z}_{2}$ symmetry and the ground
states are $\Ket{0...0}\Bra{0...0}=\Ket{\boldsymbol{0}}\Bra{\boldsymbol{0}}$
and $\Ket{1...1}\Bra{1...1}=\Ket{\boldsymbol{1}}\Bra{\boldsymbol{1}}$,
and for $h\ne0$ the symmetry bronken and the ground states depends
on the direction of the magnetic field.

The central spin interacts with the bath with an interaction strength
$\alpha\in[0,1]$ (with this interval for realistic purposes), described
using the Hamiltonian 
\begin{equation}
H_{SB}=\alpha\sigma^{z}\otimes\sum_{i}\sigma_{i}^{z},
\end{equation}
acting in the space $\mathcal{H_{SB}}\cong(\mathbb{C}^{2})^{\otimes N+1}$,
and inducing the unitary time evolution operator $U(t)=e^{-iH_{SB}t}$
- considering here the situation of interaction picture with respect
to the operator $H_{S}$; since $[H_{S},H_{SB}]=0$, arises in the
setup a case of pure decoherence, this does not affect $H_{B}+H_{SB}$
and obviously $[H_{B},\rho_{B}]=0$.

This operator $H_{SB}$ generate a quantum map $\mathcal{E}:\mathcal{L}(\mathcal{H_{S}})\rightarrow\mathcal{L}(\mathcal{H_{S}})$
and following the calculations of the Appendix \ref{ap1}, the dynamics
just affects the off-diagonal terms

\begin{equation}
\rho_{S}(t)=\begin{pmatrix}\left|a\right|^{2} & a^{*}b\Gamma(t)\\
ab^{*}\Gamma^{*}(t) & \left|b\right|^{2}
\end{pmatrix},\label{rho_s}
\end{equation}
where $\Gamma(t)$ is given by 
\begin{equation}
\begin{aligned}\Gamma(t) & =\frac{1}{Z_{B}}\sum_{\sigma}e^{-\beta E(\boldsymbol{\chi})}e^{-2it\alpha\sum_{i}\sigma_{i}}\\
 & =\frac{Z_{B}(h-2i\alpha t/\beta)}{Z_{B}(h)},
\end{aligned}
\end{equation}
and the partition function $Z_{B}$ can be compute using the transfer
matrix formalism, as described by Ref. \citep{peierls_1936}. This
function $\Gamma(t)$ is the so-called decoherence function, and,
for the present case, is a periodic function with period $\tau=2\pi/4\alpha$.

Here, we have a situation with finite time reversibility implied by
oscillations presented in the density operator coherences and, the
irreversible process is obtained by taking the continous limit \citep{lidar2020lecture},
e.g., considering each mode corresponding to a state $\boldsymbol{\chi}=(\chi_{1},...,\chi_{N})$
and defining a mode density operator $\Omega(\boldsymbol{\chi})$,
i.e., 
\begin{equation}
\Omega(\boldsymbol{\chi})=\int d\chi_{1}...\int d\chi_{N}\prod_{i=1}^{N}D(\chi_{i})\delta\left(m(\boldsymbol{\chi})-\sum_{i=1}^{N}\sigma_{i}\right),
\end{equation}
where $D(\chi_{i})$ the density of states. The partition function
can be written as 
\begin{equation}
Z_{B}=\int d\boldsymbol{\chi\,}\Omega(\boldsymbol{\chi})e^{-\beta E(\boldsymbol{\chi})}.
\end{equation}
For this case, one can speak of a decoherence rate $\gamma(t)=\log\frac{1}{\Gamma(t)}$
that describes how fast the coherence vanishes.

\subsection{Decoherence behavior and Lee-Yang zeros}

As a result, the discrete environment gives a dephasing process when
the coherence is recovered periodically that, as we will show, is
a signature of non-Markovianity. Decoherence theory provides an archetypal
mechanism for open quantum systems, which can be summarized simply
as follows: interaction between the system and the environment causes
decoherence and this, in turn, causes the loss of information from
the system to the environment \citep{Schlosshauer_2005}. Here, a
case of pure decoherence, when there is no energy dissipation in the
system, for all practical purposes, it means that we have a situation
without any effect in the population of the central qubit, and the
effect of bath interaction in the system recover elastic scattering.

The decoherence function recalls the Loschmidt amplitude, a fundamental
object of the theory of dynamical phase transitions (DQPT) \citep{Heyl_2018}.
The Loschmidt amplitude quantifies an overlap between an initial state
and its post quench evolution, meaning that this amplitude measure
how a quantum system differs from its initial state after applied
an evolution operation. One can define the Loschmidt amplitude as
\begin{equation}
\mathcal{G}(t)=\braket{\Psi_{0}|\Psi_{t}}=\Bra{\Psi_{0}}e^{-iHt}\Ket{\Psi_{0}},
\end{equation}
for some initial state $\Psi_{0}$ and a general driven Hamiltonian
$H$. Notice that the Loschmidt amplitude vanishes for the case when
the states are orthogonal. Analogous to thermal phase transitions,
DQPT's occurs when $t=t_{c}$ if $\mathcal{G}(t_{c})=0$ which results
in nonanalyticity of $\log\mathcal{G}(t)$, a dynamical analog of
the thermal free energy.

Accordingly, Loschmidt amplitude have a closed relation with the partition
function, which can be seen considering the boundary partition function
\citep{leclair1995boundary}, represented in the following form 
\begin{equation}
Z=\Bra{\Psi_{1}}e^{-RH}\Ket{\Psi_{2}},
\end{equation}
in which the states $\Ket{\Psi_{1}}$ and $\Ket{\Psi_{2}}$ encoding
the boundary conditions and $H$ denoting the bulk Hamiltonian, $R$
is the distance between two borders of the system, such as a situation
described by our system: a qubit with two energy levels corresponding
to the energy boundaries coupled to another system with a coupling
strength $\alpha$. The coupling of with respect of the qubit is described
by the interacting Hamiltonian, which can be re-writen as $\alpha(\Ket{0}\Bra{0}-\Ket{1}\Bra{1})\otimes\sum_{i}\sigma_{i}^{z}$,
and each level of the qubit is submitted to an amount $\alpha$ of
energy with respect to the bath.

Thus, we can speak of an effective Hamiltonian that computes this
behavior of the system, $H(\alpha)=2\alpha\sum_{i}\sigma_{i}^{z}$,
introducing the overlap effect caused by the thermal bath in the central
qubit, and the bath thermal state $\hat{\rho}_{B}$, we can use the
generalization for the mixed state Loschmidt amplitude given by Refs.
\citep{PhysRevB.96.180303} and \citep{PhysRevB.96.180304} to obtain
\begin{equation}
\mathcal{G}(t)=\mathrm{Tr}\left[\rho_{B}\exp\left(-2i\alpha t\sum_{i}\sigma_{i}^{z}\right)\right]=\frac{\mathrm{Tr}[e^{-\beta H_{B}}e^{-iH(\alpha)t}]}{Z_{B}},
\end{equation}
i.e., the same as decoherence function $\Gamma(t)$. In the same way,
one can consider another way to writte the coherence function with
$\Gamma(t)=\left\langle e^{-iH(\alpha)t}\right\rangle _{B}$ in which
$\left\langle \bullet\right\rangle _{B}=\mathrm{Tr}[\bullet\rho_{B}]$
denotes the thermal average with respect to the bath.

The critical times on the Loschmidt amplitude reveal a closed relation
with the zeros of the partition function, the so-called Lee-Yang zeros.
For the case of equilibrium phase transitions, the theory of Lee-Yang
tells us that the zeros of the partition function determine critical
points in the fugacity plan. The decomposition of a partition function
in the $N$th order polynomial of $z\equiv e^{-2\beta h}$ can be
obtained by 
\begin{equation}
Z_{B}=\mathrm{Tr}[e^{-\beta H_{B}}]=e^{\beta Nh}\sum_{n=0}^{N}p_{n}z^{n},
\end{equation}
where $p_{n}$ is the partition function with zero magnetic field
in which $n\le N$ spins are in the state $-1$ and $N$ the number
of spins.

The $N$ zeros of partition function lying on the unit circle in the
complex plane of $z$, can be written as $z_{n}\equiv e^{i\theta_{n}}$
with $n\in\mathbb{N}$. We re-write the partition function in function
of its zeros 
\begin{equation}
Z_{B}=p_{0}e^{\beta Nh}\prod_{n=1}^{N}(z-z_{n}).
\end{equation}
Then, the Lee-Yang zeros in the time domain, as proposed in Ref. \citep{Wei_2012},
are 
\begin{equation}
\Gamma(t)=e^{-2iN\alpha t}\frac{\prod_{n=1}^{N}(e^{-2\beta h+4i\alpha t}-z_{n})}{\prod_{n=1}^{N}(e^{-2\beta h}-z_{n})},
\end{equation}
which, of course, clarifies the one-to-one correspondence between
the decoherence function (or the Loschmidt amplitude) and the Lee-Yang
zeros. Notice that the numerator term is obtained simply by rewriting
a new (time-dependent) magnetic field $h\rightarrow h-2i\alpha t/\beta$.
When $h=0$, this function vanishes at the critical times given by
Lee-Yang zeros in fugacity plan. Then, the situation provides a setup
in which we can map an equilibrium system in a probe decoherence system.
This correspondence, beyond a mere theoretical result, guarantees
the possibility of observing Lee-Yang zeros experimentally, as can
be seen in Ref. \citep{PhysRevLett.114.010601}.

\section{\label{sec3}Non-Markovian behavior and its detection}

A crucial point for understanding the mechanism of decoherence is
the study of how information flows from the system to the environment
\citep{Schlosshauer_2005}. However, this is only part the story,
because information can also flow in the opposite direction, that
is, from the environment to the system. We call this non-Markovianity.
While in a Markovian process the open system continuously loses information
to the environment, a non-Markovian process can be characterized as
a flow of information from the environment back into the open system
\citep{rivas2014quantum,vacchini2011markovianity,li2018concepts}.

As we mentioned before, important recent contributions have been made
in terms of obtaining definitions that mean quantum counterparts for
non-Markovianity. Classically, Markovianity is reflected in the divisibility
of conditional probabilities of a stochastic process as described
by Chapmann-Kolmogorov equation \citep{VANKAMPEN200773}. Quantumly,
a definition characterized in the divisibility of quantum channels
cannot simply be imported. The propose of Rivas, Huelga, and Plenio
(RHP) \citep{rivas2014quantum} can be seen as a most similar to the
classical concept because consider that a quantum process $\mathcal{E}(t,t_{0})$
is Markovian if it is a CP-divisible map, i.e., a trace-preserving,
completely positive (CPTP) such that, for any intermediate time, it
can be divisible into two CPTP maps 
\begin{equation}
\begin{aligned}\mathcal{E}(t,t_{0})=\mathcal{E}(t,t_{1})\mathcal{E}(t_{1},t_{0}),\hspace{1em} & t_{0}\le t_{1}\le t.\end{aligned}
\label{comp}
\end{equation}
This composition between the operators frames a family of trace-preserving
and completely positive maps - a semigroup with respect to Eq.(\ref{comp}).
Any dynamics that is Markovian according to the semigroup definition
is also Markovian according to the divisibility definition, and hence
according to the BLP definition \citep{Breuer_2016}, which will be
presented in the next section.

\subsection{Trace distance-based witness of non-Markovianity}

\begin{figure*}
\includegraphics[scale=0.5]{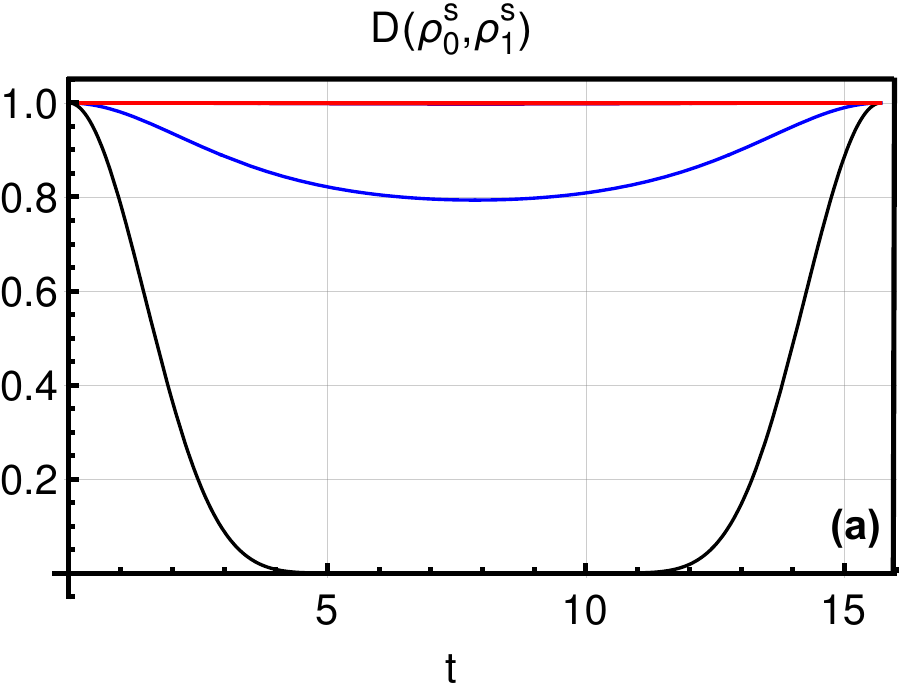}\includegraphics[scale=0.5]{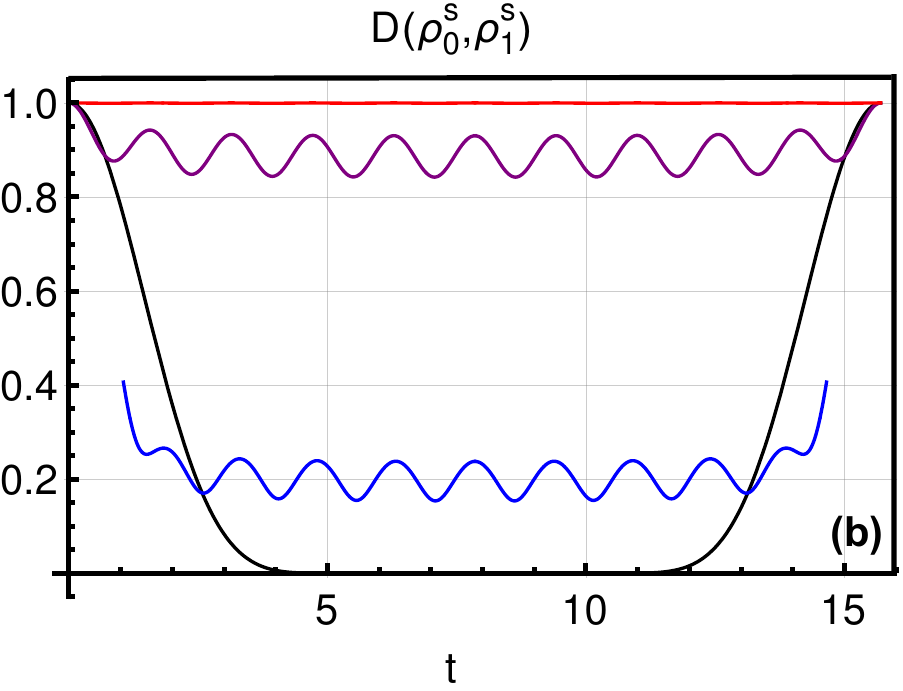}\includegraphics[scale=0.5]{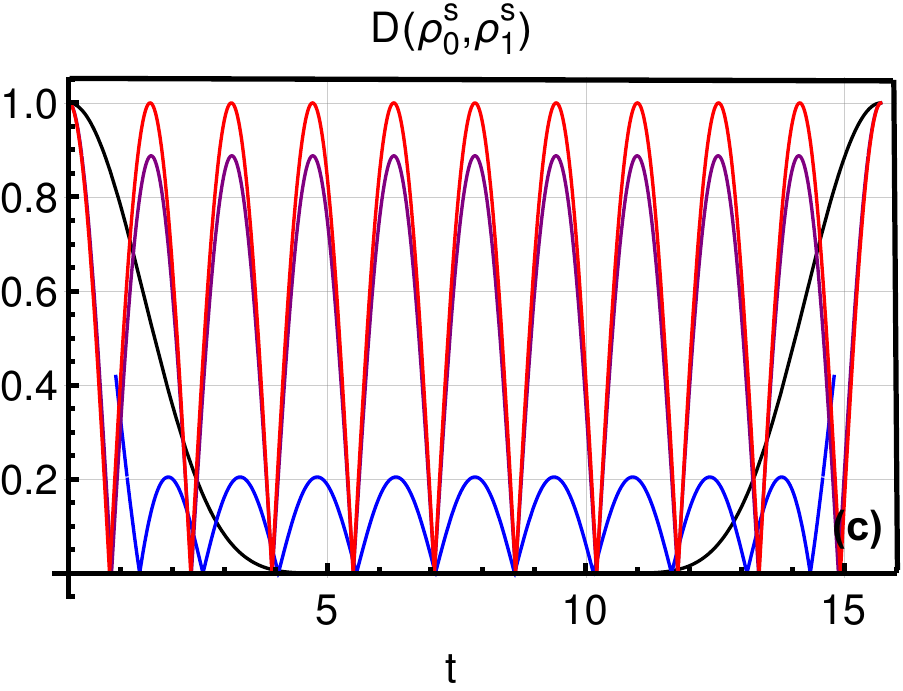}\includegraphics[scale=0.5]{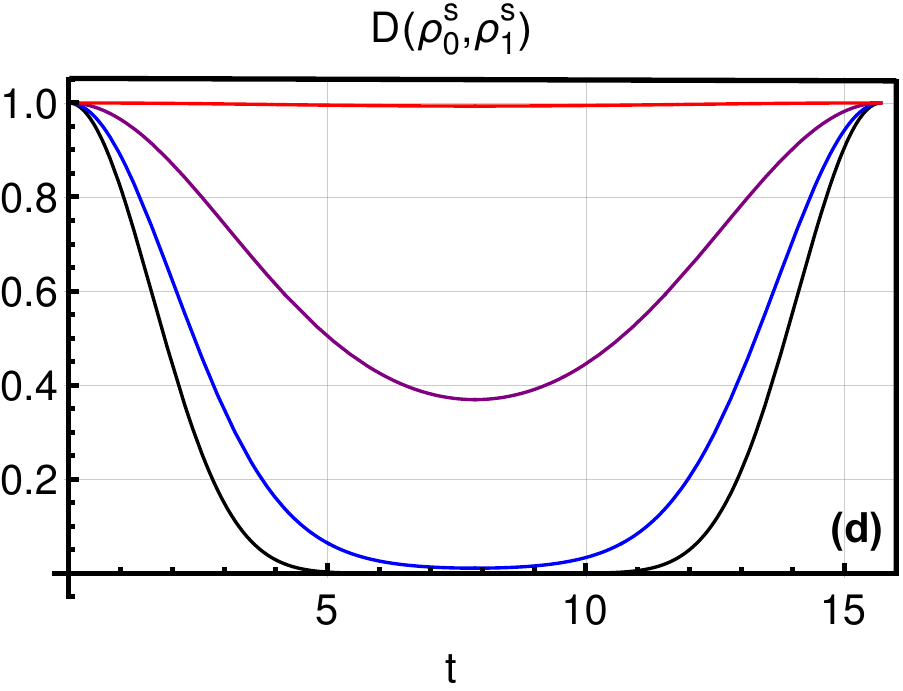}\caption{(Color online) Trace distance for two initial states $\rho_{1}=\Ket{+}\Bra{+}$
and $\rho_{2}=\Ket{-}\Bra{-}$ subject to the dynamics given by Eq.(\ref{rho_s})
for different temperatures. For the case of zero magnetic field $\textbf{(c)}$,
the trace distance oscillations correspond directly to the Lee-Yang
zeros, and this correspondence is erased as the magnetic field $h$
increases in intensity (that corresponds to $\textbf{(a)}$ and $\textbf{(b)}$).
For all cases, we see that the revivals in time culminates in situations
in which $\sigma(\rho_{1},\rho_{2};t)>0$, which would, in principle,
indicate the non-Markovianity of the system, unless for those cases
where the trace distance tends to remain constant $D(\rho_{1},\rho_{2};t)=1$
for any $t$, which cover the low temperature cases (large $\beta$).
For situations $\textbf{(a)}$, $\textbf{(b)}$ and $\textbf{(d)}$
one have $D(\rho_{1},\rho_{2})=1$ at any time for $\beta=4$ (red
line), a situation without memory effects. We set a coupling $\alpha=0.1$,
what result in the recoherence at $t=\tau$. \label{tdist}}
\end{figure*}

Markovian processes are in practice memoryless processes, i.e., processes
where information is monotonically decreasing. With this in mind,
an interesting way to obtain an intuitive and consistent definition
can be constructed by characterizing the Markovianity from distance
measures in Hilbert space \citep{laine2010measure}. The definition
of non-Markovian dynamics proposed by Breuer, Laine, and Piilo (BLP)
take into account the behavior of trace-distance. First, we will be
giving a definition of this specific Markovian condition and later
explain this meaning:

\emph{Definition. (Markovianity - BLP) An quantum evolution is Markovianity
if the distance norm between any two states decreases monotonically
with time: 
\begin{equation}
\frac{d}{dt}\left\lVert \rho_{1}(t)-\rho_{2}(t)\right\rVert _{1}\le0,
\end{equation}
where $\rho(t)=\mathcal{E}(\rho)$ and $\left\lVert X\right\rVert _{1}=\mathrm{Tr}\sqrt{XX^{\dagger}}$
is the so-called 1-norm, a measure on Hilbert space of density operators. }

The trace distance measures the indistinguishability of two states,
or the capacity to discriminate between two states. Then, the trace
distance decreases monotonically for the situations where the system
just lost information, e.g., Markovian dynamics. As a result, an increase
in its value indicates that some information is flowing back to the
system and breaks the memoryless property, a natural consequence of
the non-Markovian dynamics. Mathematically the trace distance is defined
by 
\begin{equation}
D(\rho_{1},\rho_{2})=\frac{1}{2}\left\lVert \rho_{1}-\rho_{2}\right\rVert _{1}.
\end{equation}
where we using the 1-norm defined before. Then, considering two states
evolving in time ($\rho_{1}(t)=\mathcal{E}(\rho_{1})$ and $\rho_{2}(t)=\mathcal{E}(\rho_{2})$),
it is immediate that the rate change of trace distance is given by
its first derivative, i.e., 
\begin{equation}
\sigma(t)=\frac{d}{dt}D(\rho_{1}(t),\rho_{2}(t)),
\end{equation}
and, for some $t\in[0,\infty)$, the dynamics is called non-Markovian
if 
\begin{equation}
\sigma(t)\ge0.
\end{equation}
Then, from a concise definition we have an indicator for non-Markovian
dynamics in general physical systems, where the trace distance can
be well defined. Such witness of non-Markovianity is widely used in
the literature.

In the present problem, the situation is characterized by a qubit
dephasing when put it in contact to a thermal bath. Considering two
initial pure states 
\begin{align*}
\rho_{1} & =\Ket{+}\Bra{+}, & \rho_{2} & =\Ket{-}\Bra{-}.
\end{align*}
where $|\pm\rangle=(|0\rangle\pm|1\rangle)/2$. The pure decoherence
while keep the population terms unchanged, subject the off-diagonal
terms to a modulation given by the decoherence function for the model
considered here. This case results in an immediate dependence of the
trace distance with the decoherence function, which can be written
as 
\begin{equation}
D(\rho_{1},\rho_{2};t)=|\Gamma(t)|.
\end{equation}
Then, we find that 
\begin{equation}
\frac{dD(\rho_{1},\rho_{2};t)}{dt}=\frac{\Gamma(t)}{|\Gamma(t)|}\Gamma'(t)=\mathrm{sign}[\Gamma(t)]\Gamma'(t),
\end{equation}
where $\mathrm{sign}[z]\coloneqq z/|z|$ with $z\in\mathbb{C}_{\ne0}$.
As a result, a decrease in the trace distance mean a non-Markovian
behavior as showed in the Fig. \ref{tdist}.

\begin{figure*}
\includegraphics[scale=0.3]{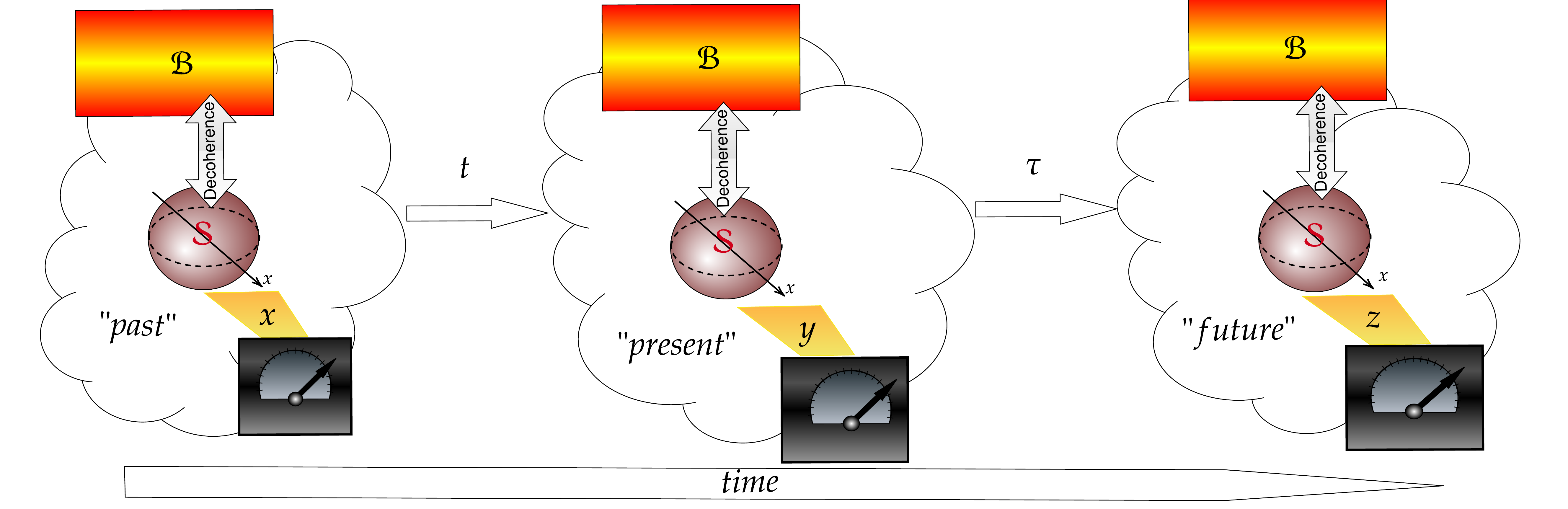}

\caption{(Color online) The conditional past-future correlator ($C_{pf}$)
is based on measurements made on the system - which, in the present
case, interacts with a thermal bath - at a time $t$ earlier and a
time $\tau$ later in relation to a present moment. The measured correlations
are related to the variables $x$, $y$, and $z$ corresponding to
the past, present, and future, respectively. Systems with strong temporal
correlations retain memory of previous states, an intuitive indication
of non-Markovianity. \label{cpfdiag}}
\end{figure*}

To understand Fig. \ref{tdist} for the trace distance, we can analyze
two different cases, where we have the interacting ($J=1$) and noninteracting
($J=0$) regimes. Considering two function defined by $C(t)\coloneqq\cosh(\beta h-2i\alpha t)$
and $S(t)\coloneqq[e^{-4\beta J}+\sinh^{2}(\beta h-2i\alpha t)]^{1/2}$,
the decoherence function for the interaction regime ($J>0$) takes
the form 
\begin{equation}
\Gamma(t)=\frac{(C(t)+S(t))^{N}+(C(t)-S(t))^{N}}{(C(0)+S(0))^{N}+(C(0)-S(0))^{N}}.
\end{equation}
To search the zeros of this function, one can consider the Lee-Yang
zeros, explicity given by the formula 
\begin{equation}
\begin{aligned}z_{n} & =-e^{-4\beta J}+(1-e^{-4\beta J})\cos k_{n}\pm\\
 & \pm\sqrt{(e^{-4\beta J}-1)\left[\sin^{2}k_{n}+e^{-4\beta J}\left(1+\cos k_{n}\right)^{2}\right],}
\end{aligned}
\end{equation}
with $k_{n}=\pi(2n-1)/N$ and $n\in\mathbb{N}$, and hence, take a
transformation in magnetic field $h\rightarrow h-2i\alpha t/\beta$
the zeros of the decoherence function reduces to: 
\begin{equation}
t=\frac{\beta h}{2i\alpha}+\frac{1}{4i\alpha}\ln z_{n}
\end{equation}
that, to results in a real times, obviously need the conditions $h\rightarrow0$
or $\beta\rightarrow0$. Here, by rewriting the decoherence function
in terms of Lee-Yang zeros we have a clear correspondence for the
critical times. In the case where the magnetic field is null, the
decoherence function touches the time axis at an interval $t=\tau$
the same number of times that zeros of the partition function occur
in the fugacity plane, as previously shown in Ref. \citep{Wei_2012}.

For the condition $J=0$, only a zero rises from the fugacity plan
for $z_{n}=-1$. The decoherence function simply reduces to 
\begin{equation}
\Gamma(t)=\frac{C^{N}(t)}{C^{N}(0)}.
\end{equation}
Thich finally results in the following zeros for the system's decoherence
function: 
\begin{equation}
t=\frac{\beta h\pm i\pi/2}{2i\alpha},
\end{equation}
that results in real terms only for weak fields $h\rightarrow0$ or
high temperatures $\beta\rightarrow0$, as the last case.

It can be seen, therefore, that this model results in a strongly non-Markovian
environment, where for low temperatures the decoherence is not affected.
Here, the recurrence at each interval $\tau$ appears as a finite
size effect, and during short intervals, the system has its initial
information restored. This effect is crucial for the storage of information
in the environment, and we see here that the information deposited
quickly returns to the system, preventing there being accessible information
in the environment to be measured.

\subsection{Conditional past-future correlation}

An also intuitive idea, proposed in Refs. \citep{Budini_2018,Budini_2019},
to characterize a Markov process, consider random variables relatives
to temporal-spaced events $x$,$y$, and $z$, such that $(x)$ is
past and $(z)$ future event with respect to a given present state
$(y)$ (Fig. \ref{cpfdiag}). These variables represents temporal-spaced
points in which one realize consecutive measures $t_{x}$, $t_{y}$,
and $t_{z}$. We expect some sort of correlation between temporal-spaced
points of states with memory and, equivalently, memoryless states
without correlation.

The idea is similar to general cases of two-points correlation functions
- more specifically, temporal correlation functions, where we measure
correlations between two spaced time points in relation to the same
system. In probabilistic language, using Bayes rule for writing the
probability $P(z,x|y)$ for the occurrence of a state $(y)$ conditioned
by a past and a future state 
\begin{equation}
P(z,x|y)=P(z|y,x)P(x|y),
\end{equation}
which $P(\cdot|\cdot)$ is the conditional probability between two
events. This property can be corroborated through a conditional past-future
correlation, which is defined as 
\begin{align*}
C_{pf} & \equiv\left\langle O_{z}O_{x}\right\rangle _{y}-\left\langle O_{z}\right\rangle _{y}\left\langle O_{x}\right\rangle _{y}\\
 & =\sum_{x,z}[P(z,x|y)-P(z|y)P(x|y)]O_{z}O_{x}.
\end{align*}
where $O_{x,y}$ is a quantity or property related to each system
state. When $C_{pf}\ne0$, one says that the system break CPF (Conditional
Past-Future) independence. Naturally, these probability distributions
are obtained from information accessible about the system from measures.
For the present case, we are accessing from measurements the information
contained in a qubit, which could be reconstructed, e.g., using quantum
state tomography. From these considerations, we can define Markovian
dynamics using the CPF measure.

\emph{Definition. (Markovianity - Budini) A quantum system evolves
between consecutive measurement events. Its dynamics are defined as
Markovian if, for arbitrary measurement processes, it does not break
CPF independence. }

\begin{figure}[h]
\begin{minipage}[t]{0.45\columnwidth}%
\includegraphics[scale=0.45]{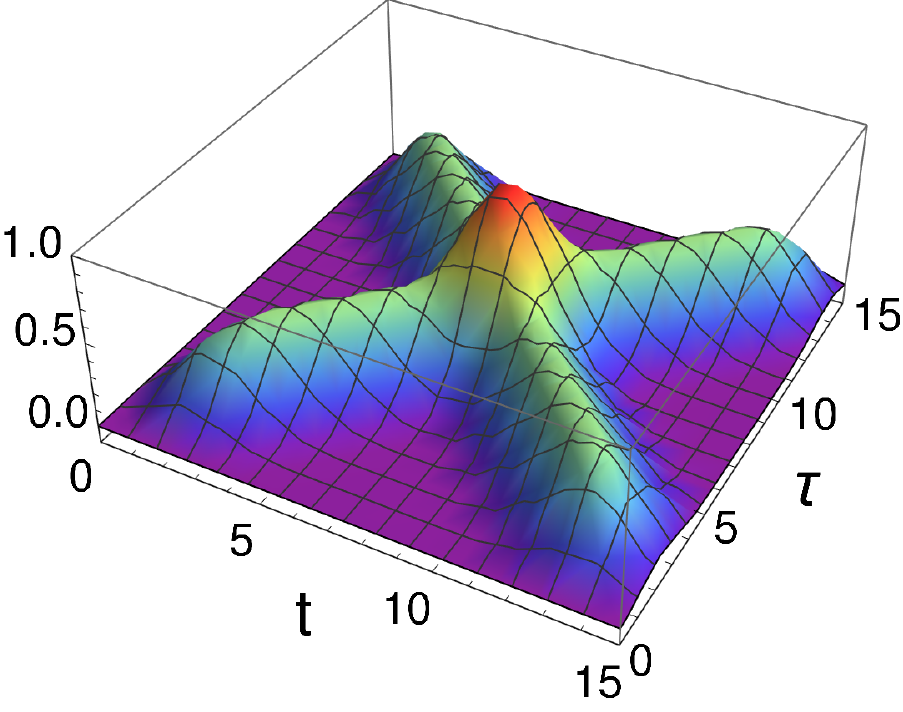}

$\textbf{(a)}$%
\end{minipage}\hfill{}%
\begin{minipage}[t]{0.45\columnwidth}%
\includegraphics[scale=0.45]{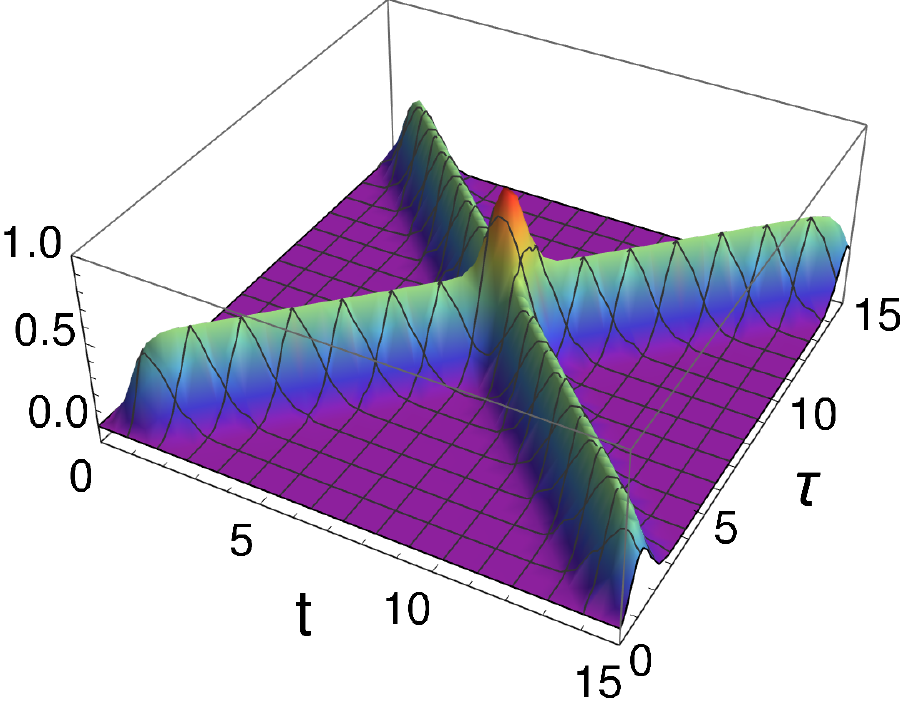}

$\textbf{(b)}$%
\end{minipage}\vfill{}
\begin{minipage}[t]{0.45\columnwidth}%
\includegraphics[scale=0.45]{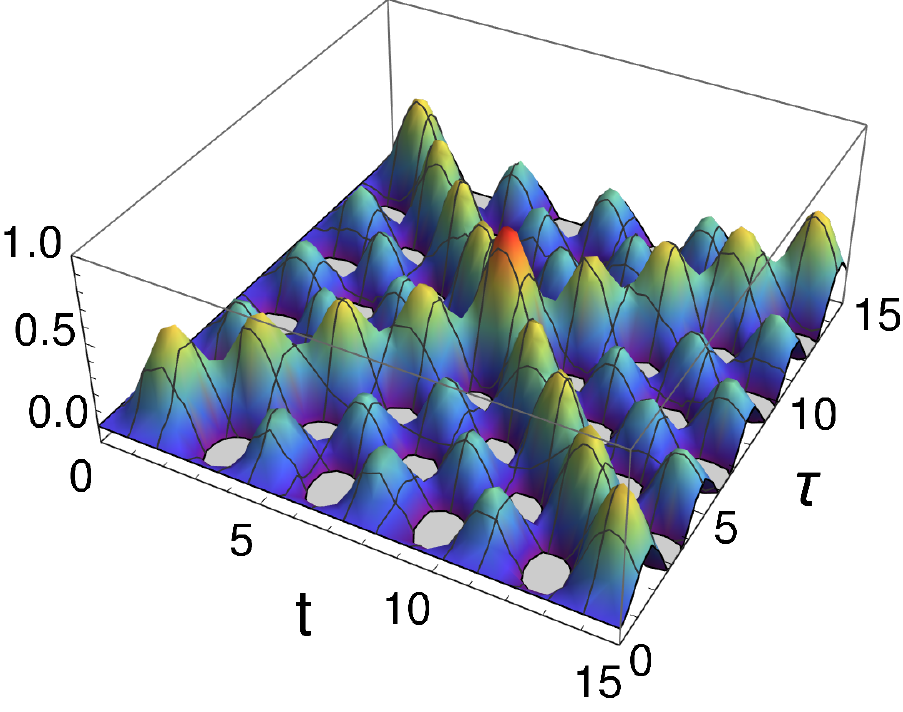}

$\textbf{(c)}$%
\end{minipage}\hfill{}%
\begin{minipage}[t]{0.45\columnwidth}%
\includegraphics[scale=0.45]{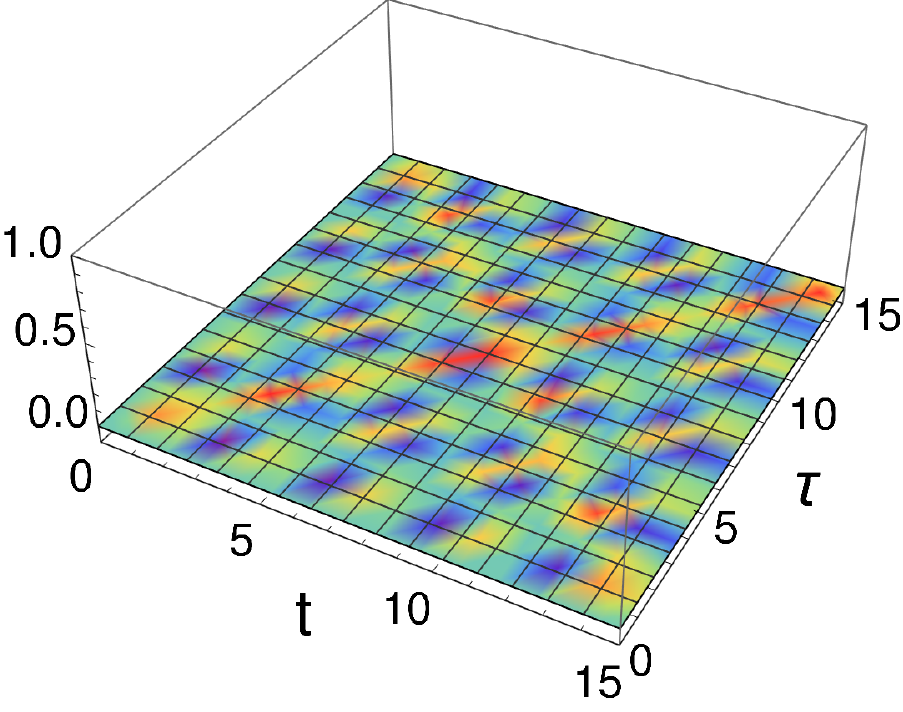}

$\textbf{(d)}$%
\end{minipage}

\caption{(Color online) Conditional past-future correlation ($C_{pf}$) for
different chain sizes and temperatures in the interacting situation
$(J=1)$. For $\textbf{(a)}$, $\textbf{(c)}$ and $\textbf{(d)}$
the chain size is $N=10$ and the temperature is $\beta=0.1$, $\beta=1$
and $\beta=4$, respectively; $N=50$ and $\beta=0.1$ in $\textbf{(b)}$.
For large $N$ and low temperatures (high $\beta$) one have system
Markovianity limits, where information is being lost to the environment.
As in trace-based measure, at $\beta=4$ the memory effects are supressed.
The situations $C_{pf}\protect\neq0$ recover situations of finite
size and temperature, which induce memory effects on the total system.
We set the coupling $\alpha=0.1$.\label{cpfi}}
\end{figure}

\begin{figure}[h]
\begin{minipage}[t]{0.45\columnwidth}%
\includegraphics[scale=0.45]{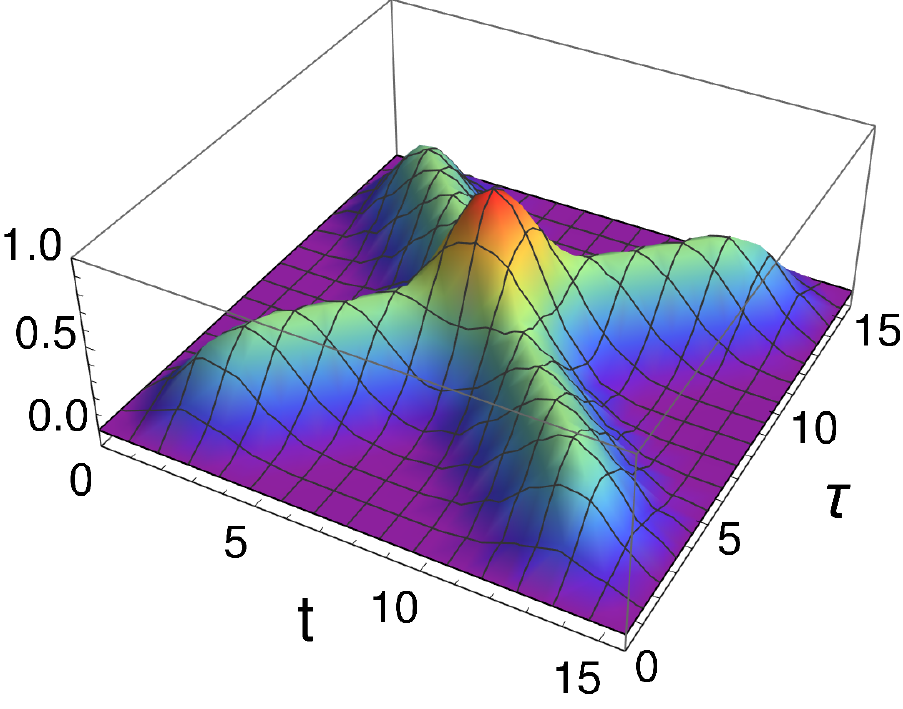}

$\textbf{(a)}$%
\end{minipage}\hfill{}%
\begin{minipage}[t]{0.45\columnwidth}%
\includegraphics[scale=0.45]{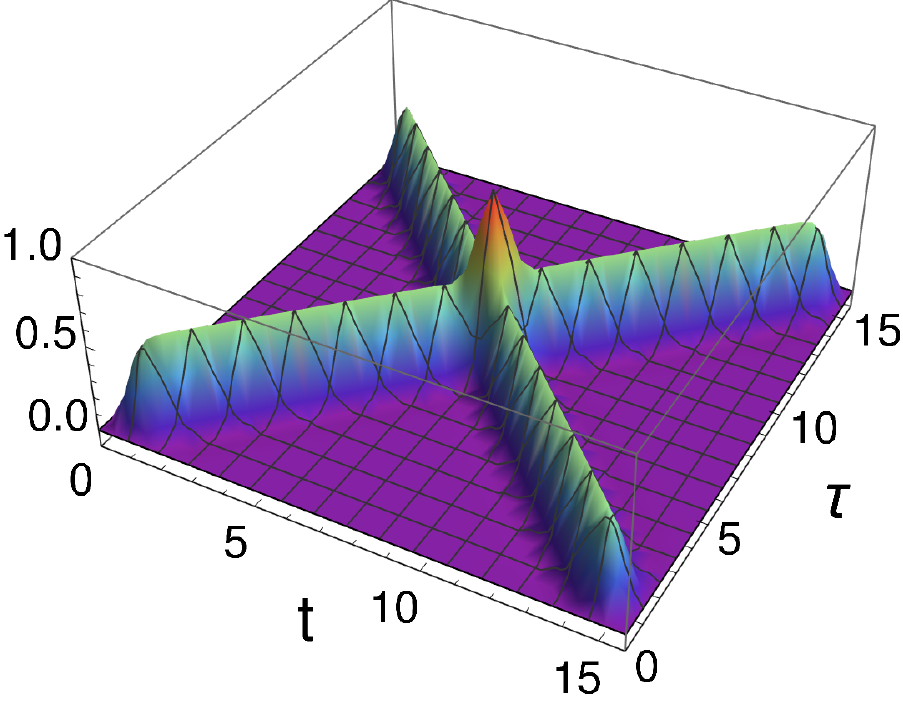}

$\textbf{(b)}$%
\end{minipage}\vfill{}
\begin{minipage}[t]{0.45\columnwidth}%
\includegraphics[scale=0.45]{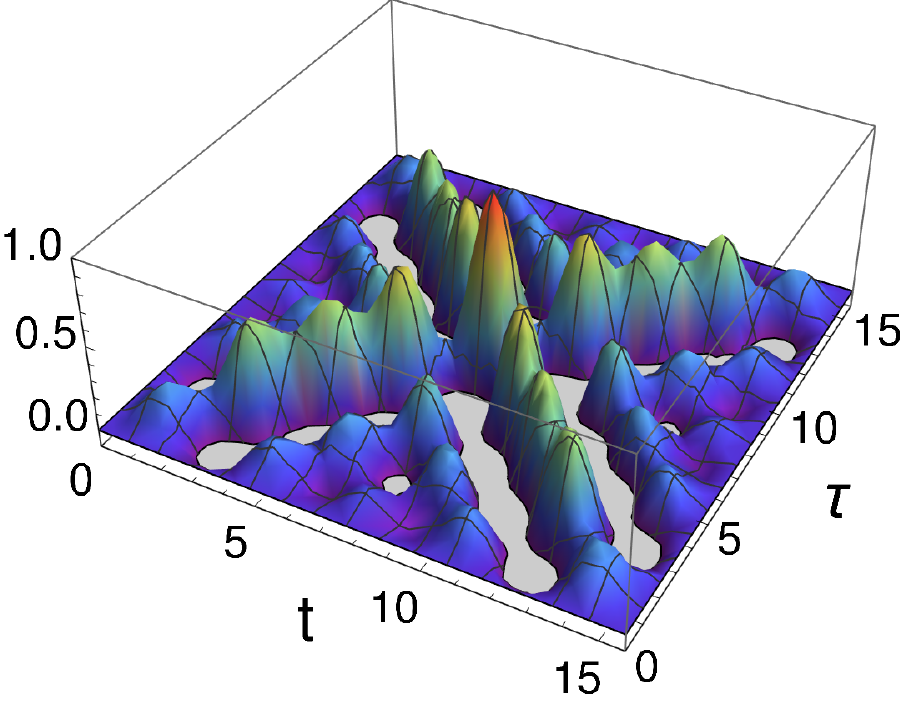}

$\textbf{(c)}$%
\end{minipage}\hfill{}%
\begin{minipage}[t]{0.45\columnwidth}%
\includegraphics[scale=0.45]{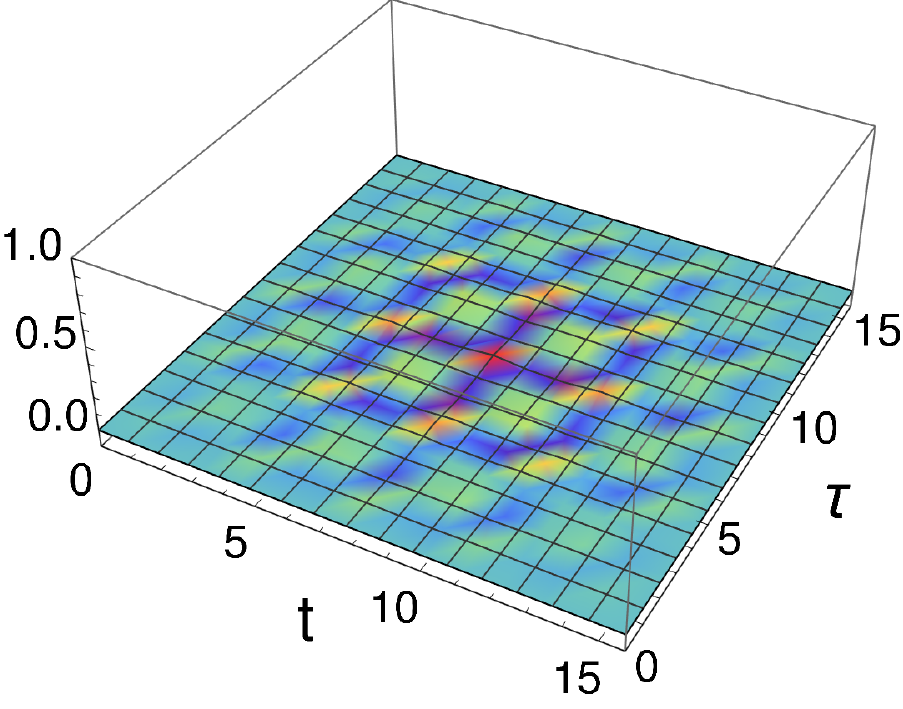}

$\textbf{(d)}$%
\end{minipage}\caption{(Color online) For $\textbf{(a)}$, $\textbf{(c)}$ and $\textbf{(d)}$
the chain size is $N=10$ and the temperature is $\beta=0.1$, $\beta=1$
and $\beta=4$, respectively; $N=50$ and $\beta=0.1$ in $\textbf{(b)}$.
The noninteracting case ($h=0$) recapture a quite similar situation,
where there is a Markovian limit for large chain sizes (large N) and
low temperature (high $\beta$). Here, we can see the same behavior
of the plot $\textbf{(d)}$ of the Fig. \ref{tdist}, correspondent
to noninteracting regime, in which the memory effects disappear for
$\beta=4$. We set the coupling $\alpha=0.1$. \label{cpf}}
\end{figure}

Follow the scheme proposed by Budini, which uses measurement operators
along the $x$ axis, i.e., $\Pi_{\pm}=\Ket{\pm}\Bra{\pm}$, with $\Pi_{\pm}$
a POVM, one can compute the conditional past-future correlation in
terms of decoherence function 
\begin{align*}
C_{pf}(t,\tau) & =\sum_{x,z}[P(z,x|y)-P(z|y)P(x|y)]O_{z}O_{x}\\
 & =f(t,\tau)-f(t)f(\tau),
\end{align*}
in which$f(t,\tau)=[f(t+\tau)+f(t-\tau)]/2$ and $f(t)=\operatorname{Re}(\Gamma(t))$.
In the present case we have a thermal bath of structured spins, and
the temperature asign new features in the system as can be seen in
Figs. \ref{cpfi} and \ref{cpf}.

\section{\label{sec4}Environment point-of-view}

When dealing with open systems, we reduce the degrees of freedom to
analyze only a portion of the system-plus-environment configuration.
However, the information that at some initial moment was contained
in the system will be deposited in an environment that, in general,
has many more degrees of freedom. So far, we have evaluated the effective
amount of the action of the environment on the states of the system,
from its perspective. But instead of having all the information about
the environment decoded in the decoherence function, we can look from
its perspective to get the behavior of the information flow between
system-bath. Then, we can summarize this session with two questions:
Is there information storage in the environment? If so, is the information
accessible from measurements? Such questions can be answered using
the paradigms of quantum Darwinism and SBS.

\begin{figure}
\includegraphics[scale=0.5]{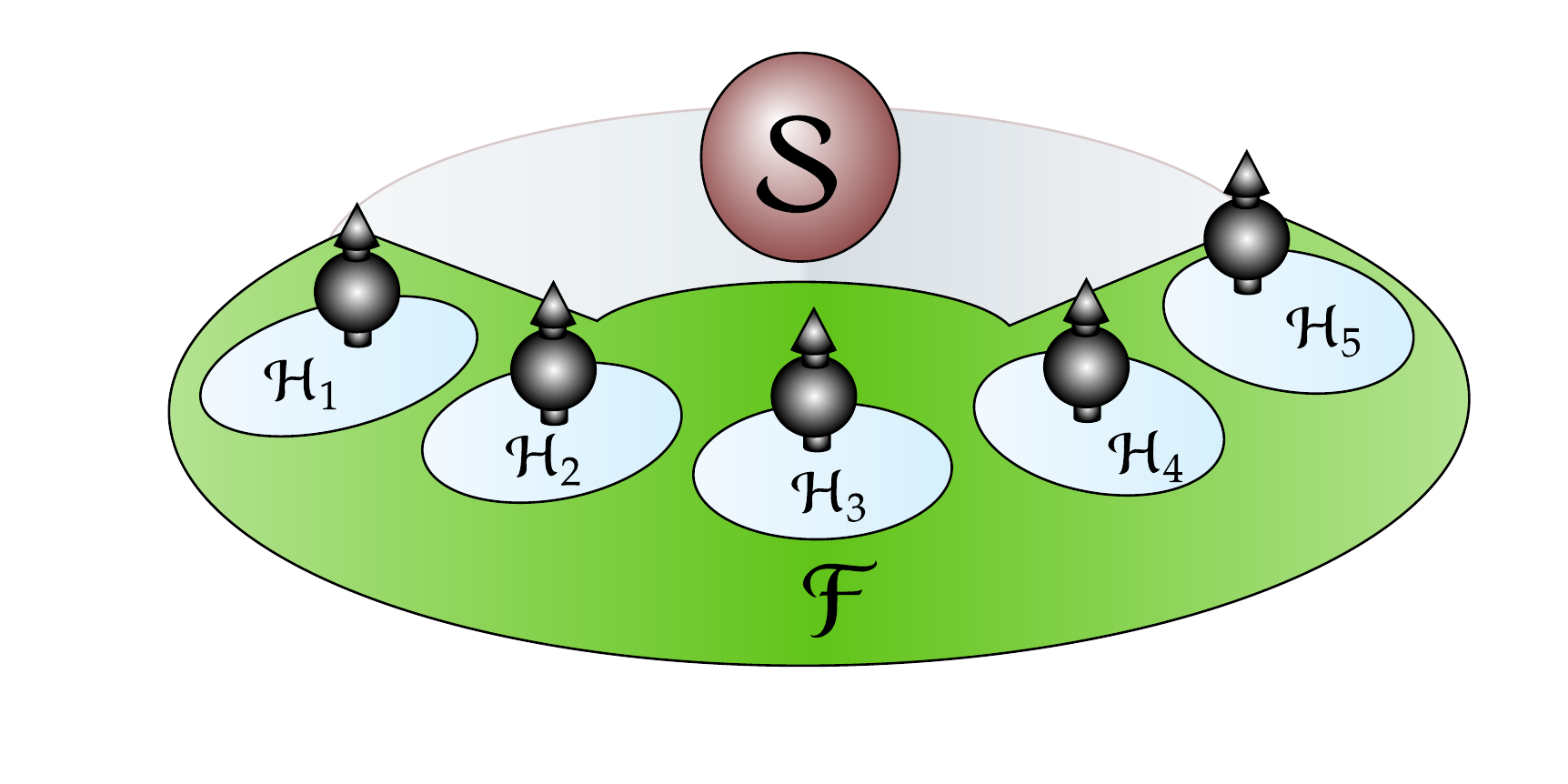}\caption{(Color online) For the description of the model in terms of the thermal
bath point of view, we use the description given in Section \ref{sec2},
in which the central qubit is described by a Hilbert space $\mathcal{H_{S}}=\mathrm{span}\left\{ \Ket{0},\Ket{1}\right\} \protect\cong\mathbb{C}^{2}$
and each environmental spin is given by $\mathcal{H}_{k}=\mathrm{span}\{\Ket{0},\Ket{1}\}\protect\cong\mathbb{C}^{2}$,
from so that the fractions are compositions of $fN$ of these spaces.\label{qdarw}}
\end{figure}

First, let us consider quantum Darwinism, an idea based on a heuristic
idea centered on the proliferation of information from a central system
to a nearby environment \citep{Zurek_2009, zurek2007relative}.
In a situation of pure decoherence, the interaction between the system
and the environment causes the destruction of superposed states. After
a certain decoherence time, one has

\begin{equation}
\rho_{S}^{dec}\approx\begin{pmatrix}\left|a\right|^{2} & 0\\
0 & \left|b\right|^{2}
\end{pmatrix}
\end{equation}
and this would be the key to understanding the emergence of classicality
in quantum systems, and the time of this process is typically extremely
short for every day, macroscale process. Being another way to analyze
the same phenomenon, quantum Darwinism makes use of a more realistic
platform: Instead of a monolithic structure for the environment, it
is divided into fractions where there is information proliferation,
and information redundancy/storage and its accessibility would be
responsible for the classical emergence, from the idea of objectivity
(Fig. \ref{qdarw}).

Roughly speaking, objectivity is the common agreement among observers
about the state of the system, which is not necessarily true for the
quantum world \citep{Roszak_2019,Korbicz2020}. Let us define the
idea more formally:

\emph{Definition. (Objectivity) A system state is objective if it
is (1) simultaneously accessible to many observers (2) who can all
determine the state independently without perturbing it and (3) all
arrive at the same result. }

Therefore, emergence of objectivity in quantum systems means emergence
of classicality. The conditions for this important link are given
recently by the framework of quantum Darwinism but critized in Ref.
\citep{Korbicz2020}, based on the importance of the possibility of
information extraction, i.e., measurable, distinguishable and accessible
information, that is not necessarily taken into account in the quantum
Darwinism paradigm.

\subsection{Quantum Darwinism}

Quantum Darwinism's approach was used to treat a wide range of systems,
like spin \citep{2021qu,Blume_Kohout_2005,Pleasance2017,riedel2012rise,ryan2020quantum},
and photonic \citep{Ciampini2018,Riedel2011,chen2019emergence,riedel2010quantum}
environments, harmonic oscillator \citep{Oliveira_2019} and Brownian
\citep{Blume-Kohout2008,Galve2016} models, and experimentally in
quantum dots \citep{Burke2010}, and photonic \citep{Ciampini2018,chen2019emergence}
setups. To study quantum Darwinism, we focus on correlations between
fragments of the environment and the system. The relevant reduced
density matrix $\rho_{S\mathcal{F}}$ is given by 
\begin{equation}
\rho_{\mathcal{SF}}=\mathrm{Tr}{}_{\mathcal{B/F}}\Ket{\psi_{SB}}\Bra{\psi_{SB}},
\end{equation}
Above, the trace is over $B$ less $\mathcal{F}$, or $B/\mathcal{F}$
- all of $B$ except for the fragment $\mathcal{F}$. Being $\mathcal{S}(\rho_{\mathcal{A}})$
the von Neumann entropy with respect to a system $\mathcal{A}$, quantum
Darwinism gives how much $\mathcal{F}$ knows about $\mathcal{S}$
can be quantified using mutual information 
\begin{equation}
I(\mathcal{S:F})=\mathcal{S}(\rho_{S})+\mathcal{S}(\rho_{\mathcal{F}})-\mathcal{S}(\rho_{\mathcal{SF}}),
\end{equation}
defined as the difference between entropies of two systems (here $\mathcal{S}$
and $\mathcal{F}$) treated separately and jointly. Thus, we can define
quantum Darwinism from the amount of shared information that proliferates
throughout the environment.

\emph{Definition. (Quantum Darwinism) There exists an environment
fraction size $f_{0}$ such that all fractions larger than it, $f\ge f_{0}$,
it holds:}

\emph{ 
\begin{equation}
\mathcal{I(S:F)}=\mathcal{S}(\rho_{S}),
\end{equation}
independently of $f$. }

\begin{figure*}
\includegraphics[scale=0.10]{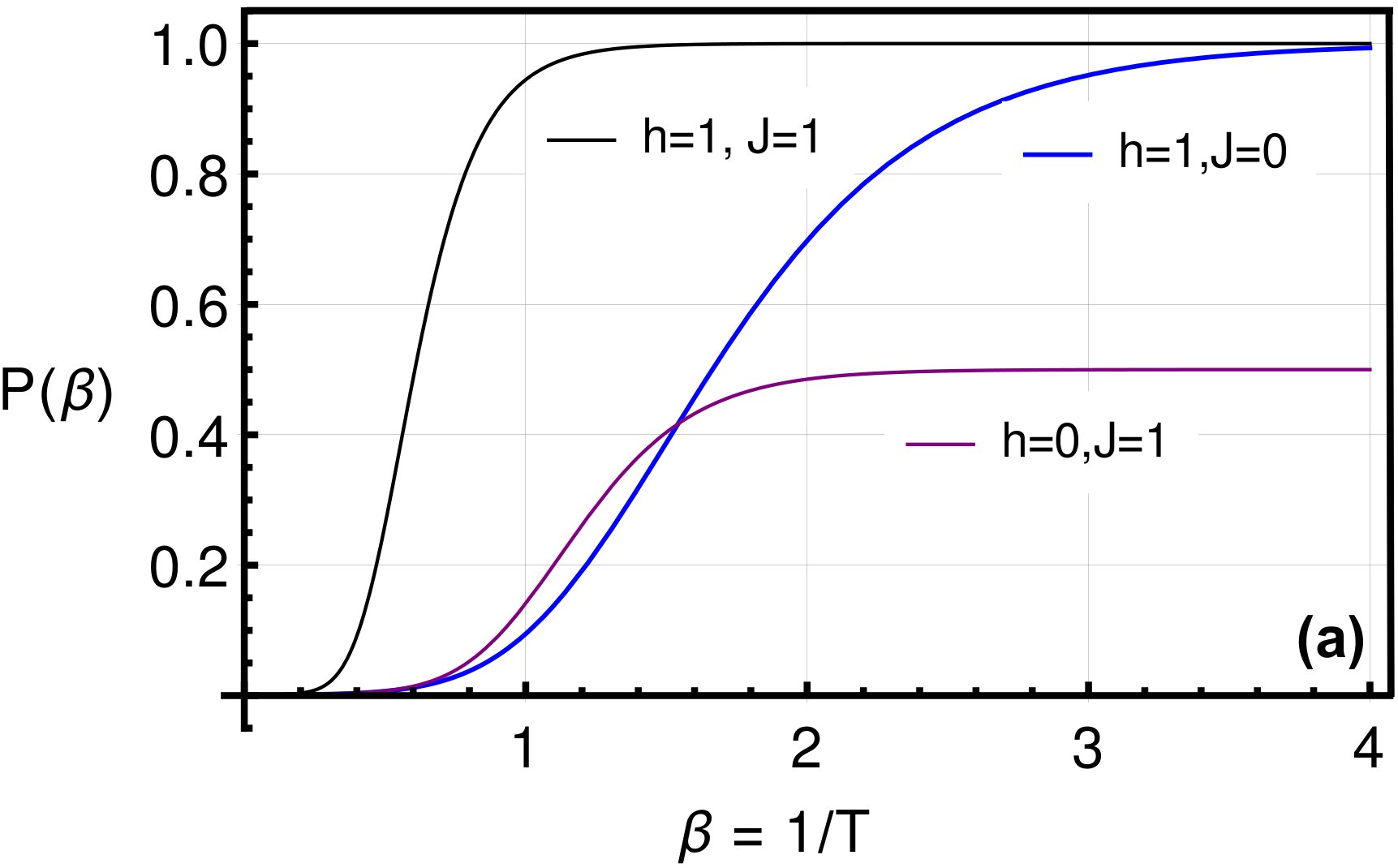}\hspace{1cm}\includegraphics[scale=0.105]{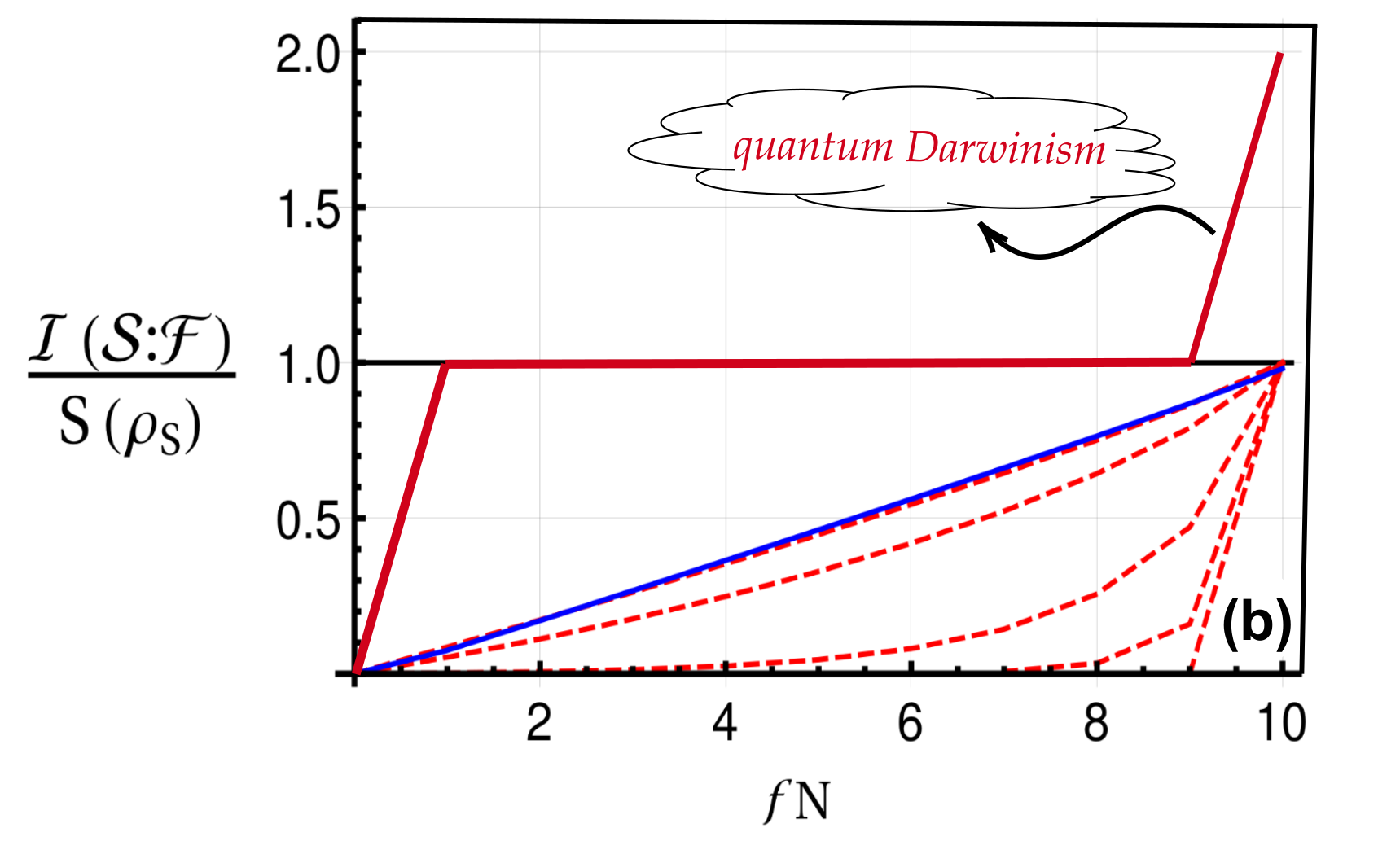}

\caption{(Color online) For the study of quantum Darwinism, we divided the
Hilbert space of the bath into fractions $\mathcal{H}_{i}\protect\cong\mathbb{C}^{2}$.
The PIPs (Partial Information Plots) method quantifies the information
between a set of $fN$ fractions and the system. The idea is to consider
that the complete knowledge of the environment about the system occurs
when the amount of correlations is $\mathcal{I(S:F)}=S(\rho_{S})$.
Here, we present (plot $\textbf{(b)}$) PIPs for $J=0$ for different
temperatures ($\beta=0.1$, $\beta=0.5$, $\beta=1$, $\beta=2$,
$\beta=4$) and an initial system state $\rho_{S}=\Ket{+}\Bra{+}$
and, respectively, $t=\tau/2$ (red dashed lines), and $t=\tau$ (blue
solid lines). For the case where $t=\tau/2$, decoherence inhibits
the storage of information in the environment more intensely as the
temperature increases. This situation sheds light on the influence
of information flow for the emergence of quantum Darwinism. In $t=\tau$,
we have total recoherence and storage is not temperature dependent
\citep{Galve2016,Milazzo2019}. Note, however, the symmetry around
f = 1/2 for the case where $\beta=4$ (and the states is pure, see
Fig. $\textbf{(a)}$) for both cases $t=\tau$ and $t=\tau/2$. For
comparison, we use the solid red line to represent an emergence of
quantum Darwinism. We set the coupling $\alpha=0.1$. \label{pipurity}}
\end{figure*}

The most direct way to check for this condition is via so-called partial
information plots (PIPs), where $\mathcal{I}(\mathcal{S}:\mathcal{F})$
is plotted as a function of $f$. The PIPs format depends on the intrinsic
characteristics of the system-environment density operator. If we
take, for example, a global pure state, we will have antisymmetric
plots around $f=1/2$ \citep{Blume_Kohout_2005,Pleasance2017,Blume-Kohout2008,Blume_Kohout_2006},
and this fact can be easily seen considering the marginal mutual informations
for the system, i.e., mutual informations correspondents to the operators
$\rho_{\mathcal{SF}}$ and $\rho_{\mathcal{S\bar{F}}}$, in which
$\mathcal{\bar{F}\coloneqq B}/\mathcal{F}$ and then 
\begin{equation}
\mathcal{I}(\mathcal{S}:\mathcal{F})+\mathcal{I}(\mathcal{S}:\mathcal{\bar{F}})=2S(\rho_{S})
\end{equation}
which stems from the fact that the marginal entropies are equal: $S(\rho_{\mathcal{SF}})=$$S(\rho_{\mathcal{S\bar{F}}})$.
But for cases where we have mixed initial states (such as thermal
states at finite temperature), such symmetry is broken in the system.
Fig. \ref{pipurity} shows the purity of the $\rho_{B}$ state as
a function of temperature for three cases (no interaction, no magnetic
field, and non-zero interaction and field).

However, a closer looking at the meaning of quantum mutual information
shows that things are not as straightforward as in classical information
theory, as we could see in the next subsection.

\subsection{Spectrum Broadcast Structures}

The focus of quantum Darwinism is on sharing information between the
system and fractions of the environment from correlations, but without
mentioning the character of these correlations or the accessibility
of information stored in the environment. For a scenario where objectivity
emerges, information about the system needs to be able to be measured.
In this way, the SBS paradigm takes into account the structure of
the states formed by the system-plus-fractions set of the environment
in order to obtain sufficient conditions for the emergence of objectivity.

SBS relates to the composition of the partially traced density operator
$\rho_{\mathcal{SF}}$. These structures have a close relation with
the possibility of quantum Darwinism and are the keys for lead the
idea of strong quantum Darwinism \citep{le2019strong,Korbicz2020}.

\emph{Definition. (Spectrum Broadcast Structure) The joint state $\rho_{\mathcal{SF}}$
of the system $\mathcal{S}$ and a collection of subenvironments $\mathcal{F}=\mathcal{E}_{1}\otimes...\otimes\mathcal{E}_{fN}$
has spectrum broadcast structure if it can be written as: 
\begin{equation}
\rho_{\mathcal{SF}}=\sum_{n}p_{n}\Ket{n}\Bra{n}\otimes\rho_{n}^{\mathcal{E}_{1}}\otimes...\otimes\rho_{n}^{\mathcal{E}_{fN}},
\end{equation}
where $\{\Ket{n}\}$ is the pointer basis of $\mathcal{S}$, $p_{i}$
are the probabilities and the operators $\rho_{n}^{\mathcal{E}_{k}}$
are perfectly distinguishable, i.e., two by two orthogonal considering
each pair of fragment environments.}

The basic idea for such structures is to consider states that can
be faithfully broadcasted satisfying Bohr non-disturbance definition:

\emph{Definition (Bohr non-disturbance) A measurement ${\Pi_{k}^{\mathcal{S}'}}$
on the subsystem $\mathcal{S}'$ is Bohr non-disturbing on the subsystem
$\mathcal{S}$ iff} 
\begin{equation}
\sum_{k}\mathds{1}\otimes\Pi_{k}^{\mathcal{S}'}\rho_{\mathcal{SS}'}\mathds{1}\otimes\Pi_{k}^{\mathcal{S}'}=\rho_{\mathcal{SS}'}
\end{equation}
Therefore, these are the states such that many observers can find
out the state $\mathcal{S}$ independently, and without perturbing
it, as assigned in the definition of objectivity. Let us assume a
system-environment interaction under the decoherence action, that
can be written as follows

\begin{equation}
\rho_{\mathcal{SF}}=\sum_{n}p_{n}\Ket{n}\Bra{n}\otimes\bigotimes_{k}\rho_{n}^{k}+\rho^{coh.},
\end{equation}
so that, the SBS structures emerges when $\rho^{coh.}=0$. The partially-traced
density operator for the $J>0$ (see appendix \ref{ap2}), 
\begin{equation}
\begin{aligned}\rho_{\mathcal{SF}}(t) & =\sum_{nm}\rho_{S}^{nm}\Ket{n}\Bra{m}\otimes\bigotimes_{i=1}^{fN}\Ket{\chi_{i}}\Bra{\chi_{i}}\times\\
 & \times\sum_{\sigma}e^{-\beta E(\boldsymbol{\chi})}e^{-i(\epsilon_{n}-\epsilon_{m})\sum_{i}\sigma_{i}t}
\end{aligned}
\end{equation}
that not express the structure that we need to broadcast all information
to environment from the system, since $e^{\pm2i\alpha m(\boldsymbol{\chi})t}\ne0$
for any time and magnetization. On the other hand, in the case of
our system, $\Gamma_{\mathcal{F}}(t)=0$ the non-interacting situation
gives us (see appendix \ref{ap3}) 
\begin{equation}
\rho_{\mathcal{SF}}(t)=\sum_{n}\rho_{S}^{nn}\Ket{n}\Bra{n}\otimes\bigotimes_{i\in\mathcal{F}}\frac{e^{\beta h\sigma_{i}^{z}}}{Z_{B}}.\label{psf}
\end{equation}
Notice that the partially-traced decoherence function can be writen
as 
\[
\Gamma_{\mathcal{F}}(t)=[\cos(2\alpha t)+i\tanh(\beta h)\sin(2\alpha t)]^{(1-f)N}.
\]
Then, that condition is only satisfied for situations in which $\beta h\rightarrow0$
and $t=\pi n/2\alpha-\pi/4\alpha$, such that $n\in\mathbb{Z}$. However,
the second condition for these states is not satisfied, because for
each pair $(n,m)$ of states with $n\ne m$, one have $\rho_{n}^{i}\parallel\rho_{m}^{i}$.
However, these are clearly not surprising situations.

As shown in \citep{Korbicz2020}, in addition to the notion of SBS
being a formalization of the emergence of objectivity in open systems,
it is also a stronger condition than quantum Darwinism for the emergence
of such. In the literature, there is also a proposal to witnessing
non-objectivity in situations of strong quantum Darwinism \citep{le2019strong}.
Strong quantum Darwinism is an extension of the theory of quantum
Darwinism that emphasizes the structure of states and their available
information \citep{Korbicz2020,Feller2021a,le2019strong}.

\section{\label{sec5}Conclusions}

For the present model, we use two distinct perspectives to assess
the information flow between the system and the bath. From the perspective
of the environment, \textcolor{black}{we used two non-Markovian witnesses,
which resulted in compatible descriptions, where the Markovian scenario
was obtained only for situations where temperatures were very small,
for different couplings between first neighbors and magnetic fields.
For the trace-distance based measure, the memory effects for $h\ne0$
vanishes at $\beta=4$ (as we show in Fig. \ref{tdist}), and the
same behavior can be obtained for the conditional past-future correlator
measure, in which $C_{pf}\approx0$ at the same temperature (see Figs.
\ref{cpfi} and \ref{cpf})}\textcolor{blue}{.} The strong non-Markovian
behavior of the bath resulted in the immediate return of information
to the system, even before the quantum recoherence period. Here, it
is worth noting that the decoherence function, in addition to having
an immediate role in the description of non-Markovianity, is also
closely linked to the Lee-Yang zeros and to the description of dynamical
phase transitions in the bath probing by central qubit: the trace-based measure vanishes same times as the appearance of zeros in the fugacity plan.

In turn, the return of information prevents the proliferation of system
states in the bath, as well as information accessible for measurement,
which returns to the system or is destroyed by the effects of temperature,
showing the fragility of quantum states subjected to thermal effects.
\textcolor{black}{For this reason, we did not obtain the appearance
of the phenomenon of quantum Darwinism, as we show in the Fig. }\ref{pipurity}\textcolor{black}{b,
as $\beta$ approaches 0, the environment has more difficult to store
the system information. Similarly, we did not see formation of SBS
in the system; even for the noninteracting situation in total decoherence
Eq. (\ref{psf}), the broadcasting states has not total distinguishability
(a condition for the formation of SBS states). Despite this, the model
studied here shows an interesting situation, in which even the degrees
of freedom of the environment being reasonably large in relation to
the system, there is a recomposition of the total information that
is lost to the environment during the dynamics.}

\section*{Acknowledgements}

The project was funded by Brazilian funding agencies CNPq (Grant No.
307028/2019-4), FAPESP (Grant No. 2017/03727-0), and by the Brazilian
National Institute of Science and Technology of Quantum Information
(INCT/IQ).

 \bibliographystyle{apsrev4-2}
 \providecommand{\noopsort}[1]{}\providecommand{\singleletter}[1]{#1}%

\appendix
\begin{widetext}
\section{\label{ap1}Exact solution}

In this section of the appendix we will calculate the exact solution
of the system density operator. Let us start with the initial density
operator of the system $\rho_{S}(0)=\Ket{\psi}\Bra{\psi}$, with $\Ket{\psi}$
as defined in Eq. (\ref{initial_state}), and fully uncorrelated with
the thermal bath described by density operator in Eq.(\ref{initial_bath}),
i.e., $\rho=\rho_{S}(0)\otimes\rho_{B}$. Then the evolved density
operator can be given by 
\begin{equation}
\rho_{S}(t)=\mathcal{E}(\rho_{S})=\mathrm{Tr}{}_{B}\left[U(t)\rho_{S}\otimes\rho_{B}U^{\dagger}(t)\right],
\end{equation}
that admits a exactly representation in the Kraus form \citep{10.5555/1972505,breuer2002theory},
i.e., 
\begin{equation}
\mathcal{E}(\rho_{S})=\sum_{\boldsymbol{\chi\chi'}}K_{\boldsymbol{\chi\chi'}}\rho_{S}K_{\boldsymbol{\chi\chi'}}^{\dagger},
\end{equation}
in which $\sum_{\boldsymbol{\chi\chi'}}K_{\boldsymbol{\chi\chi'}}^{\dagger}K_{\boldsymbol{\chi\chi'}}=\mathds{1}_{S}$
are the so-called Kraus operators. As a result, the dynamics can be
calculed writting explicitly the bath operator in energy basis, like
in Eq.(\ref{do_energy}), i.e., 
\begin{equation}
\begin{aligned}\rho_{S}(t) & =\mathrm{Tr}{}_{B}[U(t)\rho_{S}\otimes\rho_{B}U^{\dagger}(t)]\\
 & =\mathrm{Tr}{}_{B}[U(t)\rho_{S}\otimes\frac{1}{Z_{B}}\sum_{\boldsymbol{\chi}}e^{-\beta E(\boldsymbol{\chi})}\Ket{\boldsymbol{\chi}}\Bra{\boldsymbol{\chi}}U^{\dagger}(t)].
\end{aligned}
\end{equation}
Writting the trace operation explicity using a eigenbasis ${\Ket{\chi'}}\in\mathcal{{H_{B}}}$,
one can obtain the follow 
\begin{equation}
\begin{aligned} & =\sum_{\boldsymbol{\chi'}}\Bra{\boldsymbol{\chi'}}U(t)\rho_{S}\otimes\frac{1}{Z_{B}}\sum_{\boldsymbol{\chi}}e^{-\beta E(\boldsymbol{\chi})}\Ket{\boldsymbol{\chi}}\Bra{\boldsymbol{\chi}}U^{\dagger}(t)\Ket{\boldsymbol{\chi'}}\\
 & =\sum_{\boldsymbol{\chi\chi'}}\frac{e^{-\frac{\beta E(\boldsymbol{\chi})}{2}}}{Z_{B}^{\frac{1}{2}}}\Bra{\boldsymbol{\chi}}U(t)\Ket{\boldsymbol{\chi'}}\rho_{S}\frac{e^{-\frac{\beta E(\boldsymbol{\chi'})}{2}}}{Z_{B}^{\frac{1}{2}}}\Bra{\boldsymbol{\chi}}U(t)\Ket{\boldsymbol{\chi'}},
\end{aligned}
\end{equation}
where we put in a form that one can identify the Kraus operators,
that are 
\begin{equation}
K_{\boldsymbol{\chi\chi'}}=\frac{e^{-\frac{\beta E(\boldsymbol{\chi})}{2}}}{\sqrt{Z_{B}}}\Bra{\boldsymbol{\chi}}U(t)\Ket{\boldsymbol{\chi'}}.
\end{equation}
Now, let us write the evolution operator at the energy eigenbasis
to obtain the Kraus operators, i.e. 
\begin{equation}
U(t)=e^{-iH_{SB}t}=\sum_{n\boldsymbol{\chi}}e^{-it\epsilon_{n}m(\boldsymbol{\chi})}\Ket{n,\boldsymbol{\chi}}\Bra{n,\boldsymbol{\chi}},
\end{equation}
in which $m(\boldsymbol{\chi})=\sum_{i}\sigma_{i}$ is the total magnetization
spin with respect to the state $\boldsymbol{\chi}$ and $\epsilon_{n}=\alpha(-1)^{n}$
with $n=0,1$ is the energy gap obtained when one diagonalize the
operator $H_{SB}$ in the basis $\Ket{n,\boldsymbol{\chi}}=\Ket{n}\otimes\Ket{\boldsymbol{\chi}}$.
Consequently, for the Kraus operators 
\begin{equation}
\begin{aligned}K_{\boldsymbol{\chi\chi'}} & =\frac{e^{-\frac{\beta E(\boldsymbol{\chi})}{2}}}{\sqrt{Z_{B}}}\Bra{\boldsymbol{\chi}}U(t)\Ket{\boldsymbol{\chi'}}\\
 & =\frac{e^{-\frac{\beta E(\boldsymbol{\chi})}{2}}}{\sqrt{Z_{B}}}\Bra{\boldsymbol{\chi}}\sum_{n\boldsymbol{\gamma}}e^{-it\epsilon_{n}m(\boldsymbol{\gamma})}\Ket{\boldsymbol{\gamma}}\Braket{\boldsymbol{\gamma}|\boldsymbol{\chi'}}\Ket{n}\Bra{n}\\
 & =\frac{e^{-\frac{\beta E(\boldsymbol{\chi})}{2}}}{\sqrt{Z_{B}}}\sum_{n\boldsymbol{\gamma}}e^{-it\epsilon_{n}m(\boldsymbol{\gamma})}\Ket{n}\Bra{n}\delta_{\boldsymbol{\chi\gamma}}\delta_{\boldsymbol{\gamma\chi'}}\\
 & =\frac{1}{\sqrt{Z_{B}}}\sum_{n\boldsymbol{\chi'}}e^{-\frac{\beta E(\boldsymbol{\chi})}{2}}e^{-it\epsilon_{n}m(\boldsymbol{\chi'})}\Ket{n}\Bra{n}\delta_{\boldsymbol{\chi\chi'}}.
\end{aligned}
\end{equation}

Let us decompose the initial density matrix $\rho_{S}=\sum_{nm}\rho_{S}^{nm}\Ket{n}\Bra{m},$
where $\rho_{S}^{nm}=\Bra{n}\rho_{S}\Ket{m}.$ Then, aplying these
Kraus operators, an evolved state subject to the evolution take the
particular form 
\begin{equation}
\begin{aligned}\rho_{S}(t) & =\frac{1}{Z_{B}}\sum_{m,n,o,p,\chi,\chi'}\rho_{S}^{nm}e^{-\beta E(\boldsymbol{\chi})}e^{-it(\epsilon_{o}-\epsilon_{p})m(\boldsymbol{\chi'})}\times\\
 & \times\Ket{o}\underset{\delta_{on}}{\underbrace{\Braket{o,n}}}\underset{\delta_{mp}}{\underbrace{\Braket{m|p}}}\Bra{p}\delta_{\boldsymbol{\chi\chi'}}\\
 & =\frac{1}{Z_{B}}\sum_{m,n,\sigma}e^{-\beta E(\boldsymbol{\chi})}e^{-it(\epsilon_{n}-\epsilon_{m})m(\boldsymbol{\chi})}\rho_{S}^{nm}\Ket{n}\Bra{m}.
\end{aligned}
\end{equation}
and, one can easily check $\epsilon_{n}-\epsilon_{m}=\alpha[(-1)^{n}-(-1)^{m}]$
results in null terms for $n=m$, then the dynamics just affects off-diagonal
terms 
\begin{equation}
\rho_{S}(t)=\begin{pmatrix}\left|a\right|^{2} & a^{*}b\Gamma(t)\\
ab^{*}\Gamma^{*}(t) & \left|b\right|^{2}
\end{pmatrix}
\end{equation}
whilst the coherences (off-diagonal) are modulated by the periodic
function $\Gamma(t)$, given by 
\begin{equation}
\begin{aligned}\Gamma(t) & =\frac{1}{Z_{B}}\sum_{\sigma}e^{-\beta E(\boldsymbol{\chi})}e^{-2it\alpha\sum_{i}\sigma_{i}}\\
 & =\frac{Z_{B}(h-2i\alpha t/\beta)}{Z_{B}(h)},
\end{aligned}
\end{equation}
where the partition function $Z_{B}\coloneqq\mathrm{Tr}[e^{-\beta H_{B}}]$
is the Ising partition function.

\section{\label{ap2}Time evolution of the operator $\rho_{\mathcal{SF}}$
for general case}

For the present system with non-zero coupling and magnetic field,
the partially-traced density operator can be obtained expanding the
density operator $\rho_{B}$ in the energy eigenbasis (Eq. \ref{do_energy}),
in the same way as the exact solution. But, for present work, a better
path to obtain the partially-traced density operator is decompose
the time evolution operator as a tensor product, i.e., $U(t)=e^{-i\alpha\sigma^{z}\otimes\sigma_{1}^{z}t}\otimes...\otimes e^{-i\alpha\sigma^{z}\otimes\sigma_{N}^{z}t}=\bigotimes_{i=1}^{N}e^{-i\alpha\sigma^{z}\otimes\sigma_{i}^{z}t}$.
Hence, the operator comes

\begin{align*}
\rho_{\mathcal{SF}}(t) & =\mathrm{Tr}{}_{\mathcal{B/F}}[U(t)\rho_{S}\otimes\rho_{B}U^{\dagger}(t)]\\
 & =Z_{B}^{-1}\mathrm{Tr}{}_{\mathcal{B/F}}\left[\left(\bigotimes_{i=1}^{N}e^{-i\alpha\sigma^{z}\otimes\sigma_{i}^{z}t}\right)\rho_{S}\otimes\sum_{\sigma}e^{-\beta E(\boldsymbol{\sigma})}\Ket{\boldsymbol{\chi}}\Bra{\boldsymbol{\chi}}\left(\bigotimes_{i=1}^{N}e^{+i\alpha\sigma^{z}\otimes\sigma_{i}^{z}t}\right)\right],
\end{align*}
Writing each term $\Bra{n}\rho_{\mathcal{SF}}\Ket{m}\equiv\rho_{\mathcal{SF}}^{nm}$,
one have: 
\begin{align*}
\rho_{\mathcal{SF}}^{nm}(t) & =\rho_{S}^{nm}Z_{B}^{-1}\mathrm{Tr}{}_{\mathcal{B/F}}[\bigotimes_{i=1}^{N}e^{-i(\epsilon_{n}-\epsilon_{m})\sigma_{i}^{z}t}\sum_{\chi}e^{-\beta E(\boldsymbol{\chi})}\Ket{\boldsymbol{\chi}}\Bra{\boldsymbol{\chi}}]\\
 & =\rho_{S}^{nm}Z_{B}^{-1}\sum_{\boldsymbol{\chi}}e^{-\beta E(\boldsymbol{\chi})}\mathrm{Tr}{}_{\mathcal{B/F}}[\bigotimes_{i=1}^{N}e^{-i(\epsilon_{n}-\epsilon_{m})\sigma_{i}^{z}t}\Ket{\chi_{i}}\Bra{\chi_{i}}]\\
 & =\rho_{S}^{nm}Z_{B}^{-1}\sum_{\boldsymbol{\chi}}e^{-\beta E(\boldsymbol{\chi})}\bigotimes_{i=1}^{fN}e^{-i(\epsilon_{n}-\epsilon_{m})\sigma_{i}^{z}t}\Ket{\chi_{i}}\Bra{\chi_{i}}\prod_{i=fN+1}^{N}\mathrm{Tr}[e^{-i(\epsilon_{n}-\epsilon_{m})\sigma_{i}^{z}t}\Ket{\chi_{i}}\Bra{\chi_{i}}],
\end{align*}
and, finally: 
\begin{equation}
\rho_{\mathcal{SF}}(t)=\frac{1}{Z_{B}}\begin{pmatrix}|a|^{2}\sum_{\boldsymbol{\chi}}e^{-\beta E(\boldsymbol{\chi})}\bigotimes_{i=1}^{fN}\Ket{\chi_{i}}\Bra{\chi_{i}} &  & a^{\star}b\sum_{\chi}e^{-\beta E(\chi)}e^{-2i\alpha m(\boldsymbol{\chi})t}\bigotimes_{i=1}^{fN}\Ket{\chi_{i}}\Bra{\chi_{i}}\\
ab^{\star}\sum_{\boldsymbol{\chi}}e^{-\beta E(\boldsymbol{\chi})}e^{2i\alpha m(\boldsymbol{\chi})t}\bigotimes_{i=1}^{fN}\Ket{\chi_{i}}\Bra{\chi_{i}} &  & |b|^{2}\sum_{\chi}e^{-\beta E(\chi)}\bigotimes_{i=1}^{fN}\Ket{\chi_{i}}\Bra{\chi_{i}}
\end{pmatrix}
\end{equation}
that not express the structure that we need to broadcast all information
to environment from the system. Since $e^{\pm2i\alpha m(\boldsymbol{\chi})t}\ne0$
for any time and magnetization.

\section{\label{ap3}Time evolution of the operator $\hat{\rho}_{\mathcal{SF}}$
for non-interacting situation: $J=0$}

In this appendix, we show how derivate the partially reduced state
for a more simple situation (non-interacting regime). Here, the calculation
is easier by the fact of the non-interacting Hamiltonian can be describe
as $H_B=\sum_{i}H_B^i$ with $H_{B}^{i}=-h\mathds{1}_{1}\otimes...\otimes\mathds{1}_{i-1}\otimes\sigma_{i}^{z}\otimes\mathds{1}_{i+1}\otimes...\otimes\mathds{1}_{N}$.
Density operator can be re-written as 
\begin{equation}
\rho_{B}=\frac{1}{Z_{B}}\bigotimes_{i=1}^{N}e^{\beta h\sigma_{i}^{z}},
\end{equation}
in which $Z_{B}=2^{N}\cosh^{N}(\beta h)$. Of course, for Gibbs states,
noninteragent means uncorrelated. Then, one can compute each terms
of the partial reduced matrix in the follow form:

\begin{equation}
\begin{aligned}\rho_{\mathcal{SF}}^{nm}(t) & =Z_{B}^{-1}\rho_{S}^{nm}\mathrm{Tr}{}_{\mathcal{B}/\mathcal{F}}\left[\bigotimes_{i=1}^{N}e^{-i\alpha(\epsilon_{n}-\epsilon_{m})\sigma_{i}^{z}t}e^{\beta h\sigma_{i}^{z}}\right]\\
 & =Z_{B}^{-1}\rho_{S}^{nm}\mathrm{Tr}{}_{fN+1}...\mathrm{Tr}{}_{N}[e^{-i\alpha(\epsilon_{n}-\epsilon_{m})\sigma_{1}^{z}t}e^{\beta h\sigma_{1}^{z}}\otimes...\otimes e^{-i\alpha(\epsilon_{n}-\epsilon_{m})\sigma_{N}^{z}t}e^{\beta h\sigma_{N}^{z}}]\\
 & =Z_{B}^{-1}\rho_{S}^{nm}\bigotimes_{i\in\mathcal{F}}e^{-i\alpha(\epsilon_{n}-\epsilon_{m})\sigma_{i}^{z}t}e^{\beta h\sigma_{i}^{z}}\prod_{i\in\mathcal{B}/\mathcal{F}}\mathrm{Tr}\left[e^{-i\alpha(\epsilon_{n}-\epsilon_{m})\sigma_{i}^{z}t}e^{\beta h\sigma_{i}^{z}}\right],
\end{aligned}
\end{equation}
with $\rho_{\mathcal{SF}}^{nm}(t)=\Bra{n}\rho_{\mathcal{SF}}\Ket{m}$.
Explicitly writing the resulting density operator: 
\begin{equation}
\rho_{S\mathcal{F}}(t)=\frac{1}{Z_{B}}\left(\begin{array}{cc}
|a|^{2}\cosh^{(1-f)N}(\beta h)\bigotimes_{i\in\mathcal{F}}e^{\beta h\sigma_{i}^{z}} & a^{\star}b\cosh^{(1-f)N}(\beta h-2i\alpha t)\bigotimes_{i\in\mathcal{F}}e^{(\beta h-2i\alpha t)\sigma_{i}^{z}}\\
ab^{\star}\cosh^{(1-f)N}(\beta h+2i\alpha t)\bigotimes_{i\in\mathcal{F}}e^{(\beta h-2i\alpha t)\sigma_{i}^{z}} & |b|^{2}\cosh^{(1-f)N}(\beta h)\bigotimes_{i\in\mathcal{F}}e^{\beta h\sigma_{i}^{z}}
\end{array}\right)
\end{equation}

We can do the identifications that follow 
\begin{align*}
\rho_{\mathcal{F}}= & \bigotimes_{i\in\mathcal{F}}\frac{e^{\beta h\sigma_{i}^{z}}}{\mathrm{Tr}\left[e^{\beta h\sigma_{i}^{z}}\right]}, & \rho'_{\mathcal{F}}(t) & =\bigotimes_{i\in\mathcal{F}}\frac{e^{(\beta h-2i\alpha t)\sigma_{i}^{z}}}{\mathrm{Tr}\left[e^{\beta h\sigma_{i}^{z}}\right]}
\end{align*}
and the decoherence function comes 
\[
\Gamma_{\mathcal{F}}(t)=\left[\frac{\cosh(\beta h-2i\alpha t)}{\cosh(\beta h)}\right]^{(1-f)N}.
\]
Where we finally were able to explicitly write the matrix for the
partially traced density operator 
\begin{equation}
\rho_{\mathcal{SF}}(t)=\begin{pmatrix}|a|^{2}\rho_{\mathcal{F}} &  & a^{\star}b\rho'_{\mathcal{F}}(t)\Gamma_{\mathcal{F}}(t)\\
ab^{\star}\rho'_{\mathcal{F}}(t)\Gamma_{\mathcal{F}}(t) &  & |b|^{2}\rho_{\mathcal{F}}
\end{pmatrix}.
\end{equation}
In this limit the calculations for the fraction entropy can be done
easily, because the fraction density operator is a tensorial composition
of $fN$ $2\times2$ matrices.
\end{widetext}

\end{document}